\documentclass[aps,twocolumn,showpacs,showkeys,superscriptaddress,amsmath,amssymb,nofootinbib,floatfix,prc]{revtex4-1}

\usepackage{graphicx}
\usepackage{subfigure}
\usepackage{bm}

\makeatletter

\begin{document}

\title{Azimuthally sensitive femtoscopy in event-by-event hydrodynamics}

\author{Piotr Bo\.zek}
\email{Piotr.Bozek@ifj.edu.pl}
\affiliation{AGH University of Science and Technology, Faculty of Physics and Applied Computer Science, al. Mickiewicza 30, PL-30059 Krakow, Poland}
\affiliation{The H. Niewodnicza\'nski Institute of Nuclear Physics, Polish Academy of Sciences, PL-31342 Krak\'ow, Poland}

\begin{abstract}

We analyze the   pion femtoscopy correlations  in noncentral Au-Au and Pb-Pb collisions. The
 azimuthally sensitive  Hanbury Brown-Twiss (HBT) method is used to extract the 
 interferometry radii depending on the
 azimuthal  angle with respect to the second and third-order event plane. The results are in semiquantitative  agreement with the STAR collaboration data on the HBT 
radii with respect to the second-order reaction plane, with the preliminary
 PHENIX collaboration data on the HBT radii with respect to the third-order 
reaction plane in  Au-Au collisions at $200$~GeV, and with the preliminary ALICE collaboration data for the   HBT radii with respect to the second-order 
event plane for Pb-Pb 
collisions at $2.76$~TeV.
\end{abstract}

\pacs{25.75.-q, 25.75.Gz, 25.75.Ld}

\keywords{ultra-relativistic proton-nucleus collisions, relativistic 
  hydrodynamics, collective flow, HBT correlations, RHIC, LHC}

\maketitle

\section{Introduction \label{sec:intro}}

The collective expansion of the dense fireball in heavy-ion collisions 
manifests itself as a significant transverse, azimuthally asymmetric flow 
\cite{Florkowski:2010zz,*Heinz:2013th,*Gale:2013da,*Luzum:2013yya}.
The azimuthal asymmetry of the flow originates from the geometrical asymmetry
 of the participant region in the transverse plane \cite{Ollitrault:1992} and
from  event-by-event fluctuations of the fireball shape 
\cite{Alver:2006wh,Alver:2010gr,Alver:2010gr,*Petersen:2010cw,*Alver:2010dn}.

The size of the emission region can be estimated using femtoscopy methods 
based on HBT interferometry for identical particles 
\cite{Wiedemann:1999qn,*Heinz:1999rw,*Lisa:2005dd,*Lisa:2011na}. The HBT radii are  extracted from a Gaussian
fit to the  two-pion correlation function. The  
radii decrease as function of the  average transverse momentum of the pair,
following the change in the 
 size  of the homogeneity region \cite{Akkelin:1995gh}.
The radii measured at different  energies at the Relativistic 
Heavy Ion Collider (RHIC) \cite{Abelev:2009tp,Adams:2004yc}  and the Large Hadron Collider (LHC) \cite{Aamodt:2011mr}
indicate a strong correlation between the velocity of the emitting 
source and its position. The observed  correlations are
 consistent with  the existence 
of a strong 
collective flow 
\cite{Broniowski:2008vp,*Pratt:2008qv,*Bozek:2010er,*Karpenko:2012yf}.
 
Azimuthally asymmetric emission geometry and transverse flow lead
 to angle dependent correlation radii 
 \cite{Voloshin:1995mc,Wiedemann:1997cr,Lisa:2000ip,Heinz:2002sq,Heinz:2002au}.
The correlation functions are constructed for pairs  of particles 
in a bin in azimuthal angle. The radii extracted 
from pairs flying in-plane are different than for pairs flying out-of-plane.
The analysis allows to extract the dependence of the HBT radii on the 
azimuthal angle. The direction is defined with respect to the event plane.
The measured azimuthal dependence of the HBT radii shows
 a zeroth  and second-order harmonic  for Au-Au collisions  at the AGS  
\cite{Lisa:2000xj} and at RHIC \cite{Adams:2003ra}, 
 Au-Pb collisions at the SPS \cite{Adamova:2008hs},
 and Pb-Pb collisions 
at the LHC \cite{Gramling:2012xqa,logginswpcf}.

The presence of the triangular flow makes it possible to study the azimuthal dependence of the HBT radii with respect to the third-order event plane
\cite{Voloshin:2011mg}.
The PHENIX collaboration presented preliminary results
\cite{Niida:2013lia,niidawpcf} showing a third-order harmonic 
modulation of the {\it out} and {\it side} radii.
The theoretical investigation of the relation between the triangular 
flow and geometry modulations and the observed femtoscopy radii \cite{Plumberg:2013nga} concludes that the observed HBT radii modulation is determined by
 the flow. The shape triangularity at freeze-out  cannot 
be directly extracted, unlike the spatial eccentricity.

Azimuthally sensitive femtoscopy radii with respect to the second-order event plane have been extracted from hydrodynamical simulations  
\cite{Frodermann:2007ab,Kisiel:2008ws}. The results are in fair agreement 
with experiment. The description of the triangular flow and of the azimuthal
 dependence  of the HBT radii with respect to the third-order event plane 
requires the application of event-by-event simulations.
The analysis of the angle averaged femtoscopy radii in the hydrodynamic model
with  fluctuating initial conditions indicated a sizable effect of the lumpiness of the emission surface for ideal fluid evolution
  \cite{Socolowski:2004hw} and  a small effect for viscous fluid 
expansion \cite{Bozek:2012hy}. Generally, one expects a strong 
damping of anisotropies through viscous evolution \cite{Schenke:2010rr}.
Angle averaged and azimuthally sensitive HBT radii have been calculated
 for a range of energies in the ultrarelativistic quantum molecular
 dynamics model 
\cite{Li:2012ta,Graef:2013wta}.
In this paper we present the calculation of the azimuthally sensitive HBT radii for Au-Au collisions at $200$~GeV and Pb-Pb collisions at $2.76$~TeV
within a $3+1$ dimensional ($3+1$-D) viscous hydrodynamic model 
\cite{Bozek:2011ua}.

\section{The model}

\label{sec:model}

\begin{figure}
\includegraphics[angle=0,width=0.22 \textwidth]{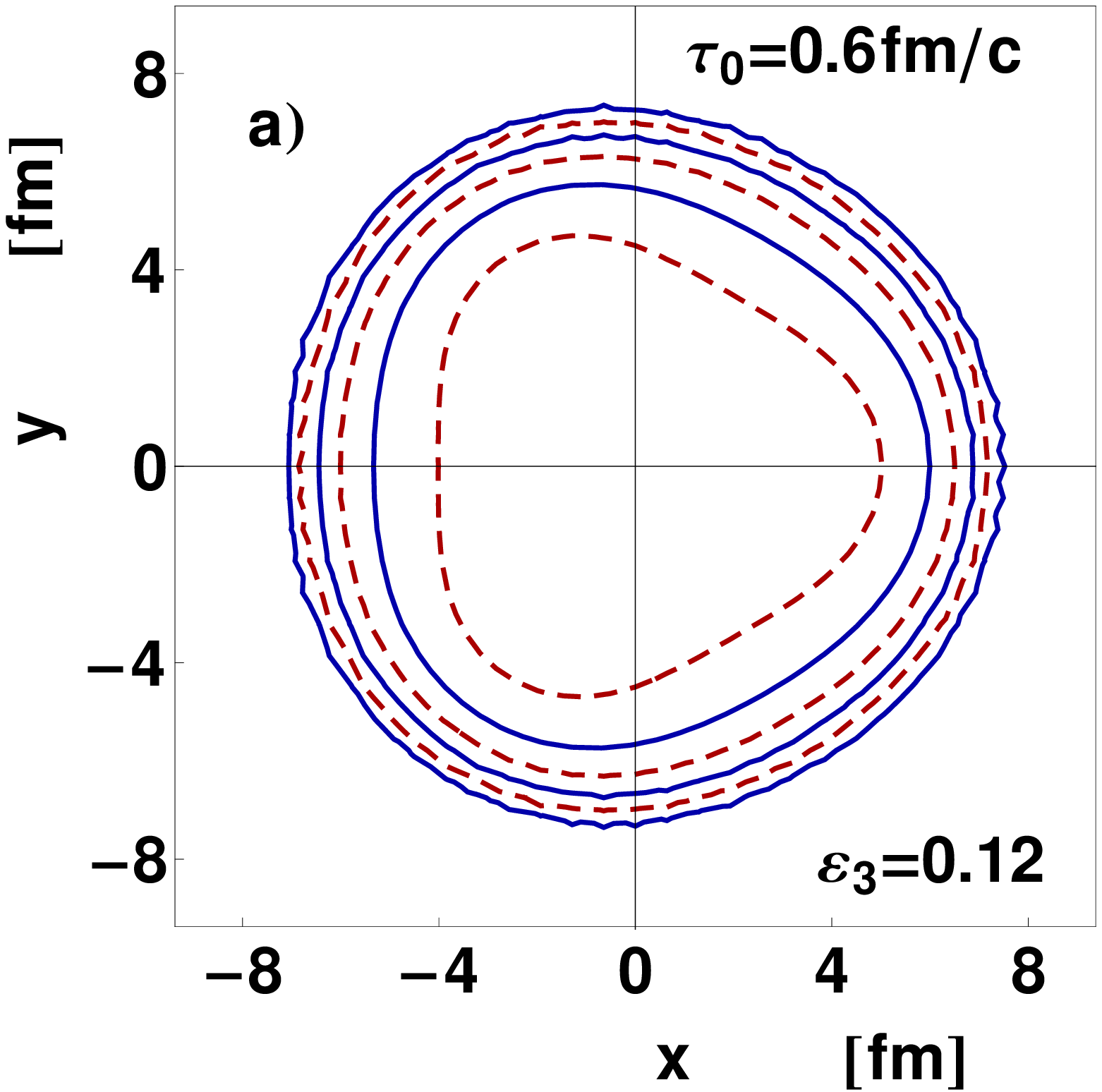} \includegraphics[angle=0,width=0.22 \textwidth]{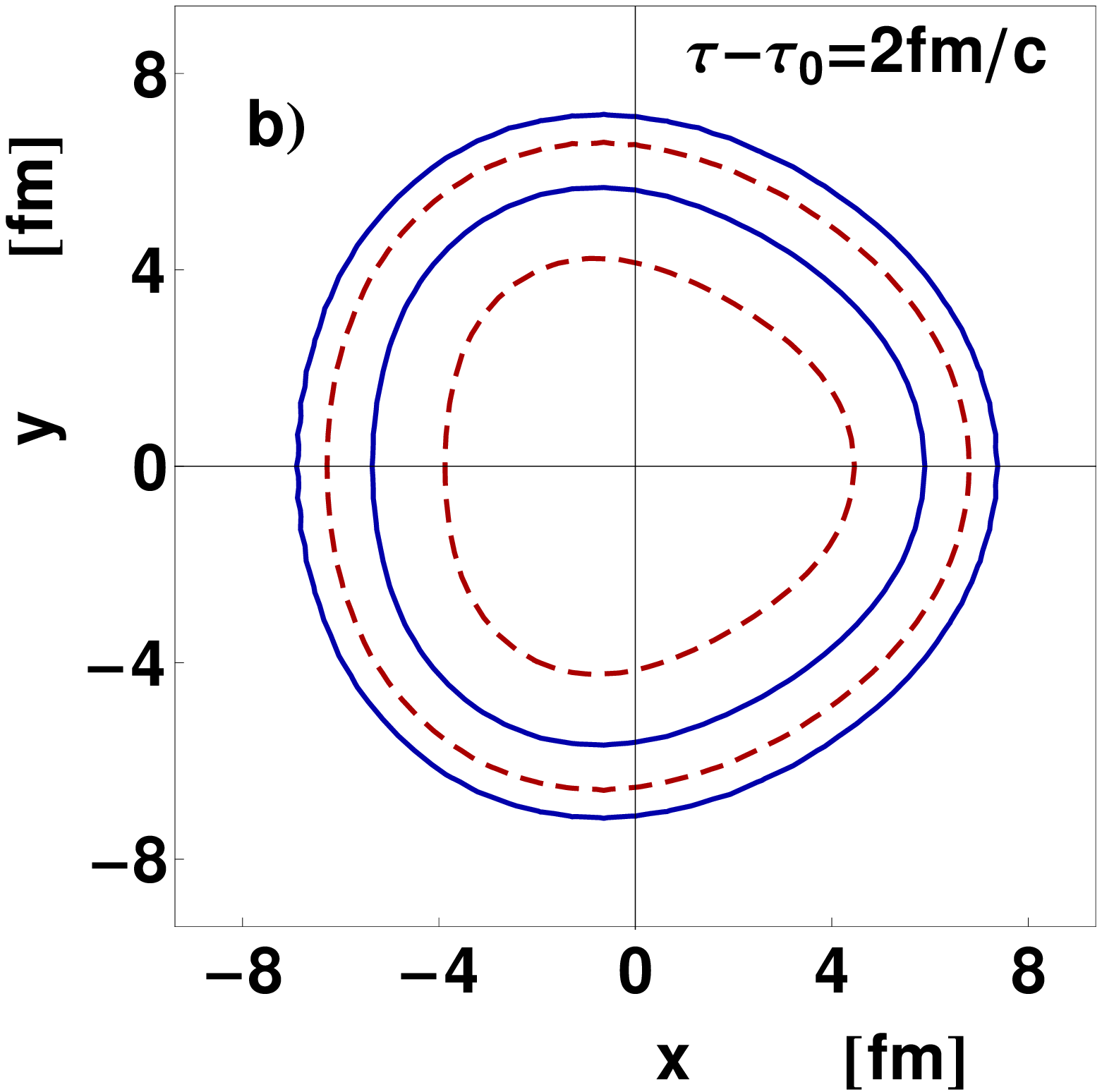}\\ 
\vskip 2mm

\includegraphics[angle=0,width=0.22 \textwidth]{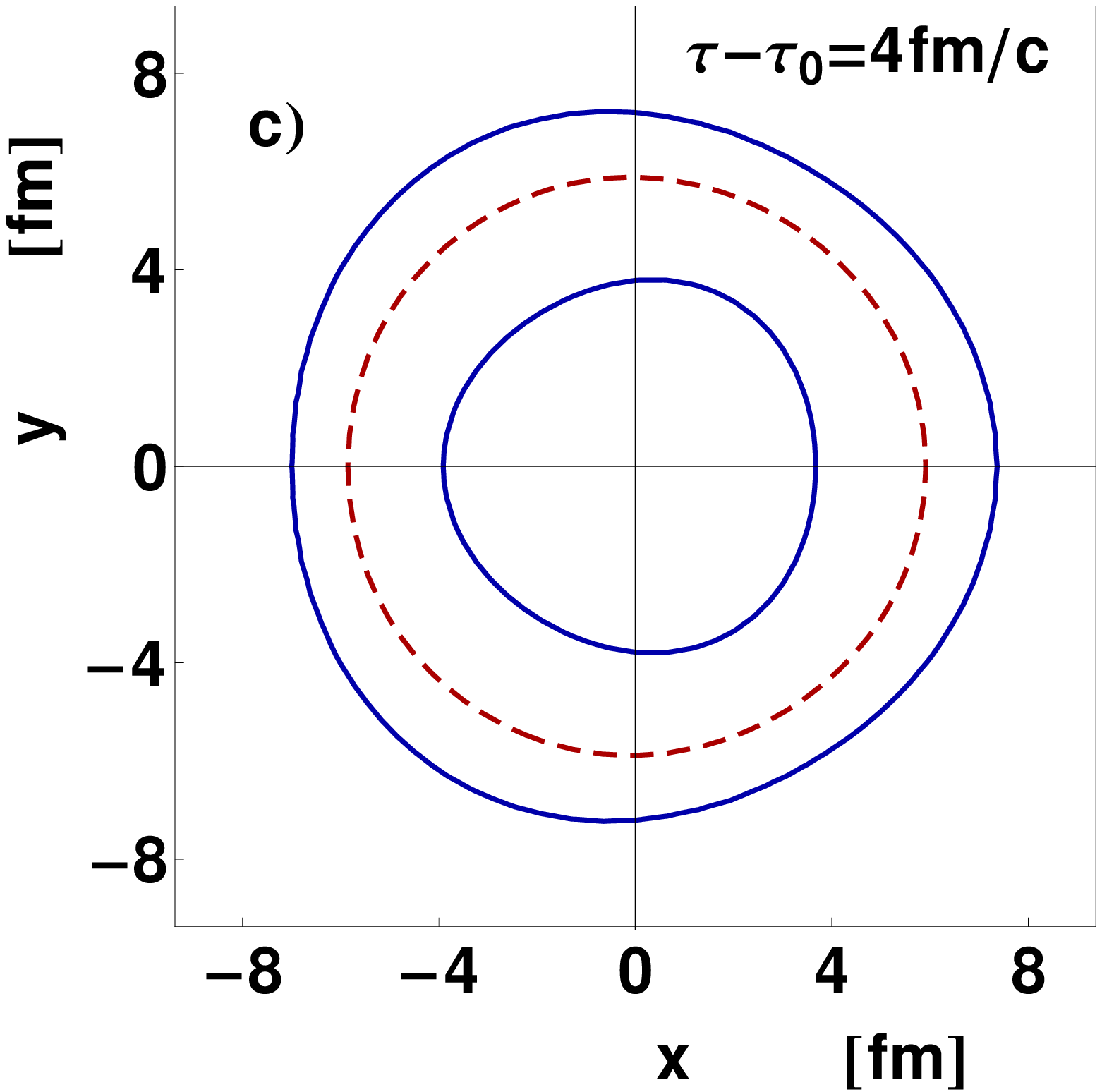} \includegraphics[angle=0,width=0.22 \textwidth]{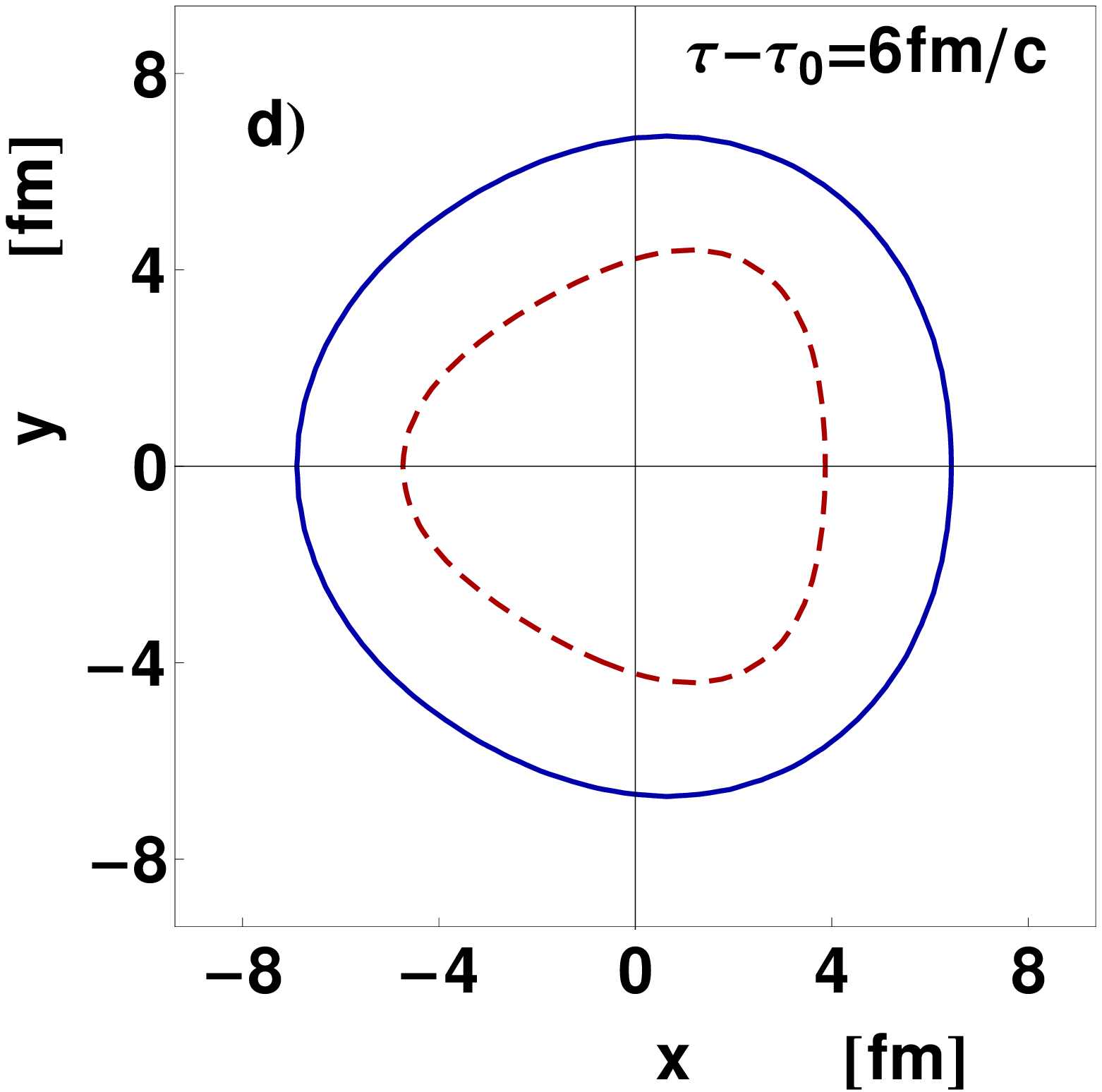}\\ 

\vskip 2mm

\includegraphics[angle=0,width=0.22 \textwidth]{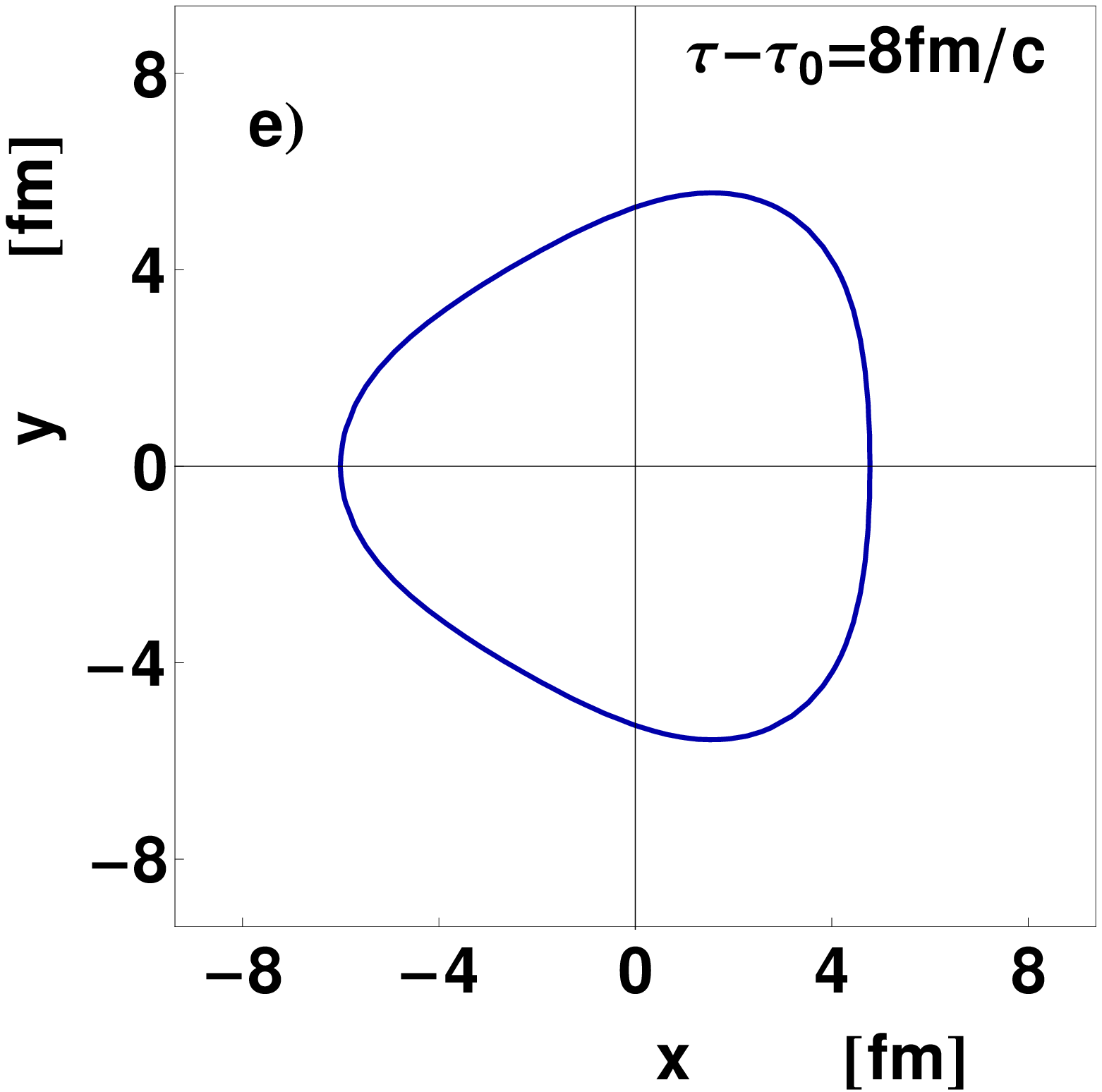}
\caption{(Color online) Contours of constant entropy density of the fireball in the transverse plane ($s=2$fm$^{-3}$, $4$fm$^{-3}$, $8$fm$^{-3}$, $16$fm$^{-3}$, \dots). The consecutive panels ( a) trough e) ) show the fireball at different fluid 
 evolution times
$\tau_0=0.6$~fm/c, $\tau-\tau_0=2$,  $4$,  $6$, and  $8$fm/c. The initial distribution
 corresponds to the optical Glauber model  density at the impact parameter $b=0$, with additional triangularity
 $\varepsilon_3=0.12$.
\label{fig:cont}} 
\end{figure}   

The expansion of the fireball is described using $3+1$-D hydrodynamics \cite{Bozek:2011ua}.
The simulations here presented use a  constant shear viscosity to entropy ratio $\eta/s=0.08$ and 
 bulk viscosity that is nonzero in the hadronic phase  $\zeta/s=0.04$ \cite{Bozek:2009dw}.
The initial density of the fireball is obtained from a Glauber Monte Carlo model 
\cite{Rybczynski:2013yba}. Every event of the Glauber Monte Carlo model constitutes an initial condition for an independent  hydrodynamic evolution.
The entropy density in the transverse plane is composed from
  Gaussians of width $0.4$~fm at the positions of the participant nucleons, details can be 
found in \cite{Bozek:2011if,Bozek:2012fw}. At the freeze-out hypersurface of constant temperature 
$T_f=150$ (or $140$~MeV when specified) hadrons are emitted statistically. The
 statistical hadronization and resonance decays are performed using the
  THERMINATOR code~\cite{Chojnacki:2011hb}. For each freeze-out hypersurface a large number of
real events are generated, by running the THERMINATOR code many times. We take from $100$ events
for central Pb-Pb collisions at $2.76$~TeV up to 1500 events for semiperipheral Au-Au
 collisions at $200$~GeV. The number of independent hydrodynamic runs for each centrality class is
$100$-$400$.

The expansion velocity grows faster in the direction of the largest gradient. This stronger flow tends
to make the spatial distribution more uniform. Depending on the length of the evolution, the
final shape of the fireball may be close to azimuthal symmetry, or in some cases the 
asymmetry could be inverted with respect to its initial direction. On  the other hand, the deformation 
of the flow
\begin{eqnarray}
&& v(r,\phi)= \nonumber \\ && v(r) (1+2 a_2 \cos\left(2(\phi -\psi_2)\right)+2 a_3 \cos\left(3(\phi -\psi_3)\right) 
\end{eqnarray}
is well correlated in its direction $\psi_n$ to the original 
direction of the elliptic or triangular deformation.  The blast-wave models \cite{Retiere:2003kf}
assume some form of the geometry and flow deformation. On the other hand,
accurate dynamical simulations can 
capture different realizations of the flow and geometry deformations at freeze-out. By comparison 
to the data, this
 can provide a phenomenological constraint on the degree of final azimuthal asymmetry of the shape.

The  change of the azimuthal asymmetry of the fireball in time can be illustrated in a 
simple example. We take as the initial entropy density the optical Glauber 
model density $\rho(x,y,b=0)$
 at the impact  parameter $b=0$, for Au-Au collisions at the top RHIC energy.
 The distribution is modified to introduce a nonzero triangularity
\begin{equation}
\rho(x,y)=\rho(x,y,b=0)e^{\frac{e_3 \cos(3\phi)r^2}{2\langle r^2\rangle}}
\label{eq:ini3}
\end{equation}
where $e_3=0.3$ and
\begin{equation}
\langle r^2 \rangle=\frac{\int \rho(x,y,b=0) (x^2+y^2) dx dy}{\int \rho(x,y,b=0)  dx dy}=17.5\ 
{\rm fm}^2 \ .
\end{equation}
The initial triangularity is $\varepsilon_3=0.12$.

The transverse  expansion is the strongest in the directions $\pm \pi/3$ 
and $\pi$ (Fig. \ref{fig:cont}).
After the evolution time of  $4$~fm/c 
 the geometrical shape is approximately symmetric. 
Further expansion reverses
 the deformation (Fig. \ref{fig:cont}, panels d) and e)).

\section{Azimuthally sensitive HBT}

\begin{figure}
\includegraphics[angle=0,width=0.32 \textwidth]{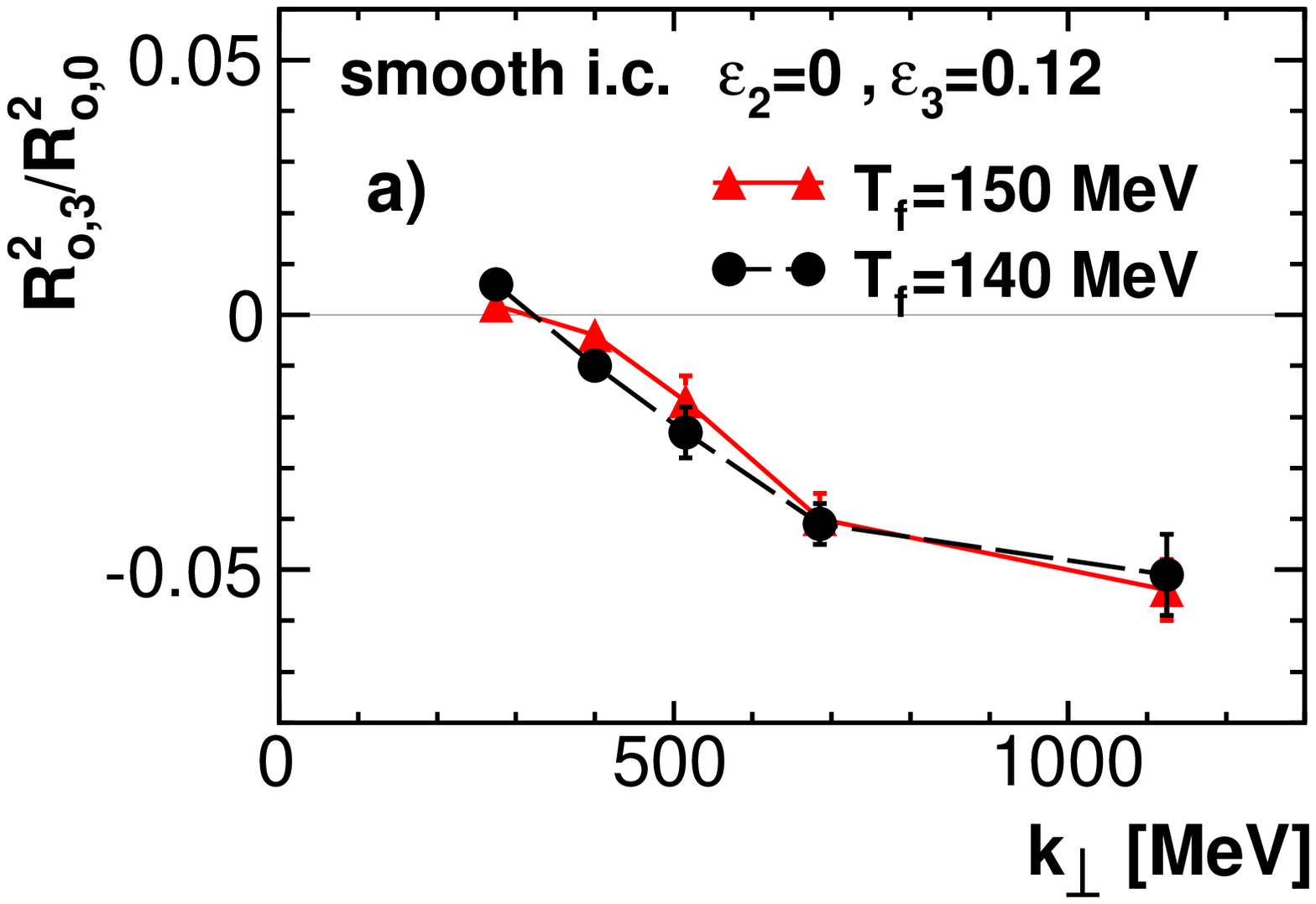} 
\includegraphics[angle=0,width=0.32 \textwidth]{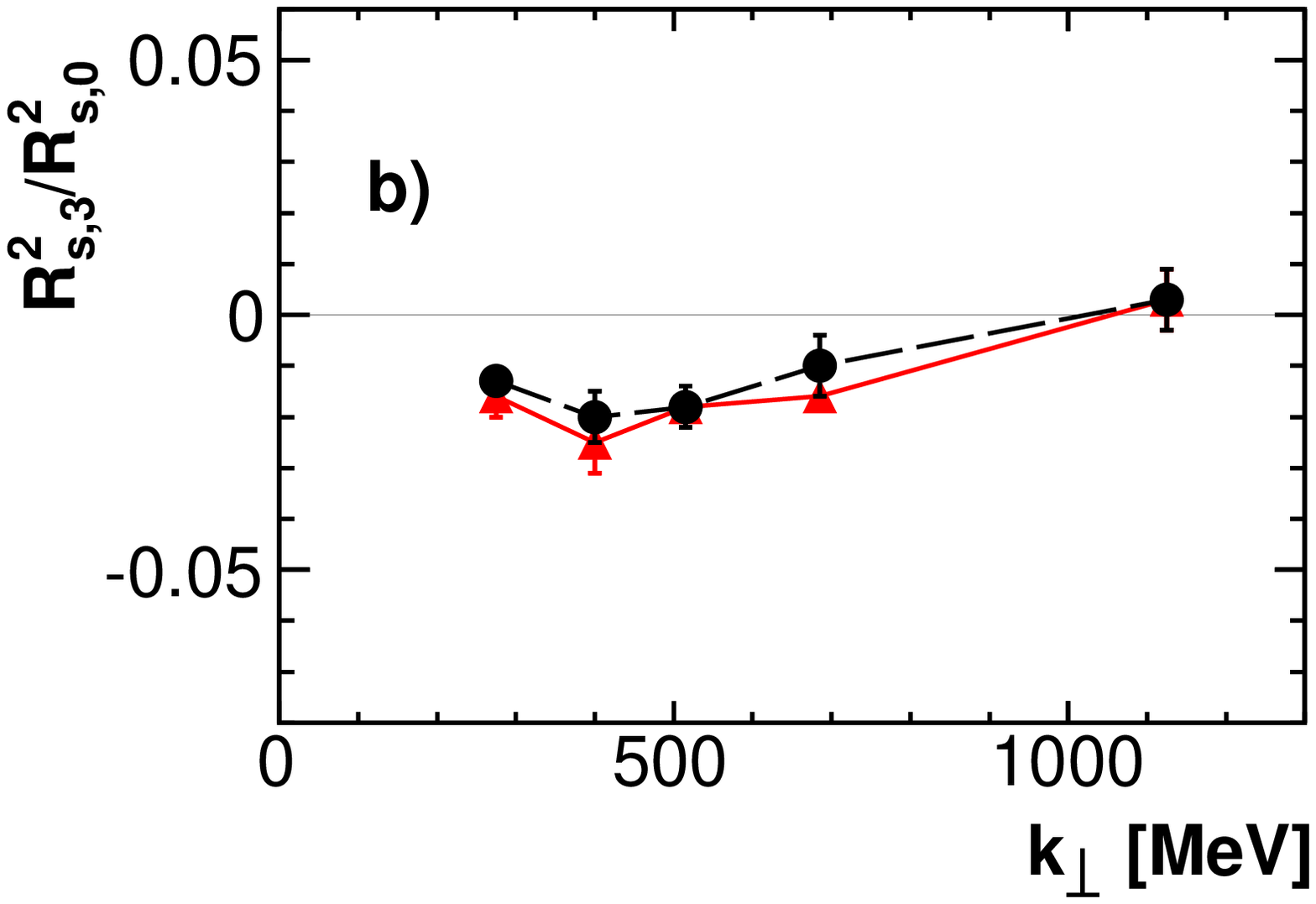} 
\includegraphics[angle=0,width=0.32 \textwidth]{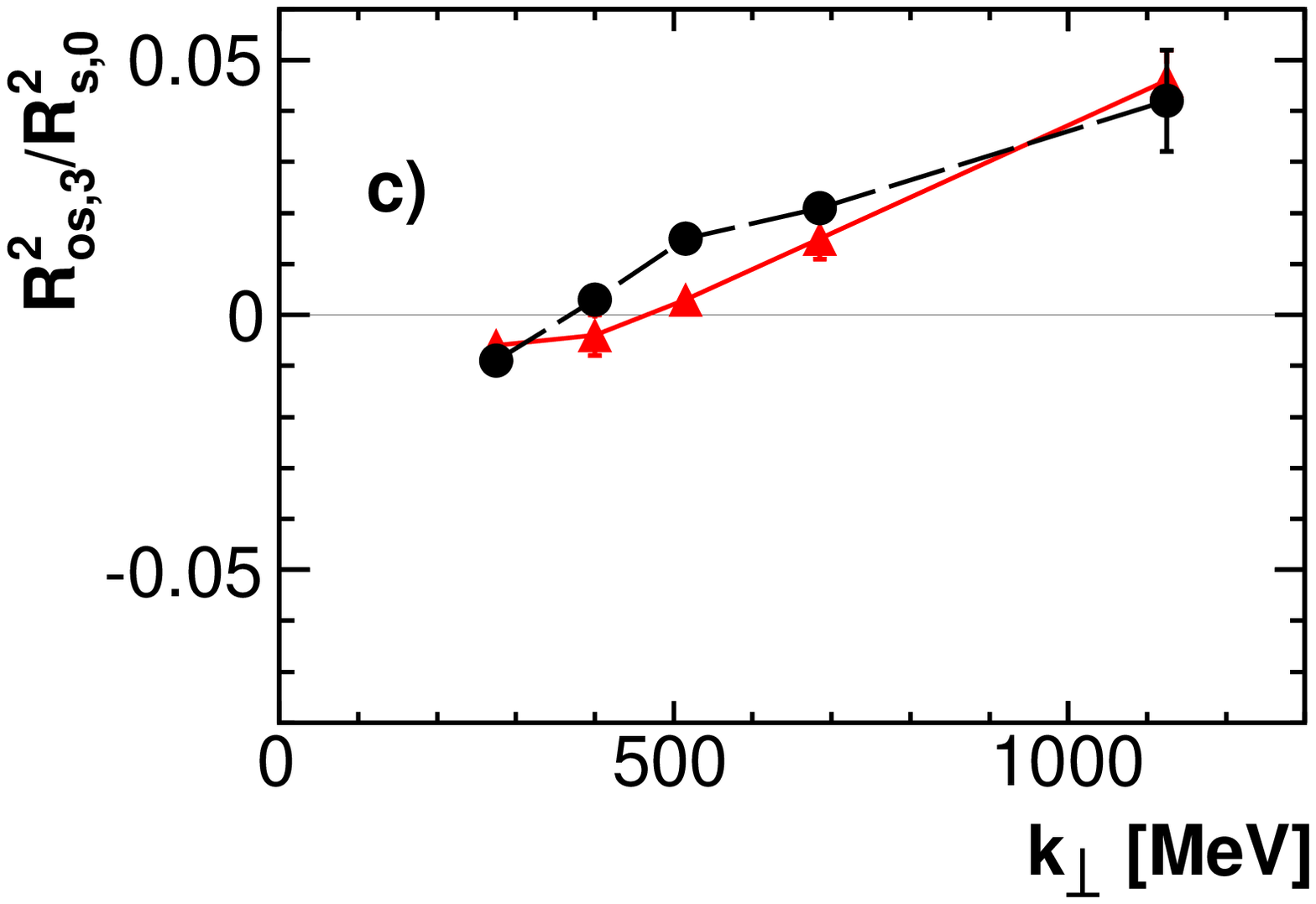} 
\caption{(Color online) Third-order Fourier coefficient of the azimuthal oscillations of the  HBT radii  with respect to third-order event plane as function of the
 transverse momentum for 
smooth initial conditions (Eq. \ref{eq:ini3}), $R_{o,3}^2/R_{o,0}^2$ (panel a), $R_{s,3}^2/R_{s,0}^2$ (panel b),   $R_{os,3}^2/R_{s,0}^2$ (panel c). 
\label{fig:model}} 
\end{figure}

The symmetrized 
correlation function of two pions  with momenta $p_{1,2}$ emitted at freeze-out from positions 
$x_{1,2}$ is
\begin{eqnarray}
 && C(q,k)= \nonumber \\
&& \frac{ \int d^4x_1 d^4x_2 \langle S(x_1,p_1)S(x_2,p_2) \rangle |\Psi(k,(x_1-x_2))|^2}
{\int d^4x_1 \langle  S(x_1,p_1)\rangle  \int d^4x_2 \langle S(x_2,p_2) \rangle} \ , \nonumber \\  
\label{eq:cq}
\end{eqnarray}
where  $q=p_1-p_2$ is the relative momentum of the pions, $k=(p_1+p_2)/2$ is the average pair momentum,
 $\Psi(q,x_1-x_2)=(e^{i q (x_1-x_2)}+ e^{-iq(x_1-x_2)})/\sqrt{2}$ is the two-particle wave function \cite{Wiedemann:1999qn}. 
We neglect
 final state interactions between the pions. In the experimental analysis  Coulomb corrections are applied 
to the measured correlation functions \cite{Bowler:1991vx,*Sinyukov:1998fc}. 
In the simulation we assume that the emitted pion do not interact, and no 
Coulomb corrections are made   on the constructed two-particle correlations. 

\begin{figure}
\includegraphics[angle=0,width=0.32 \textwidth]{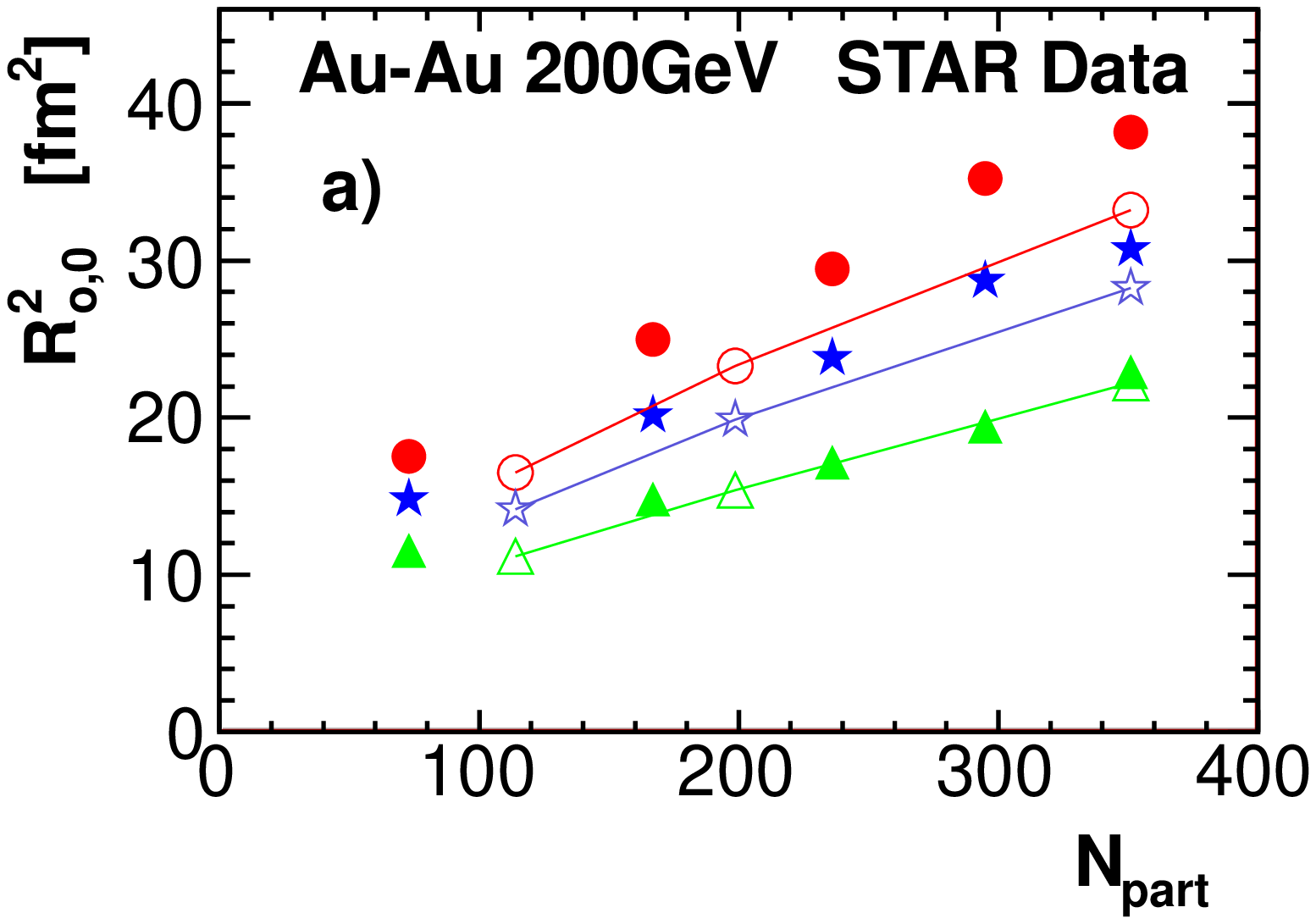} 
\includegraphics[angle=0,width=0.32 \textwidth]{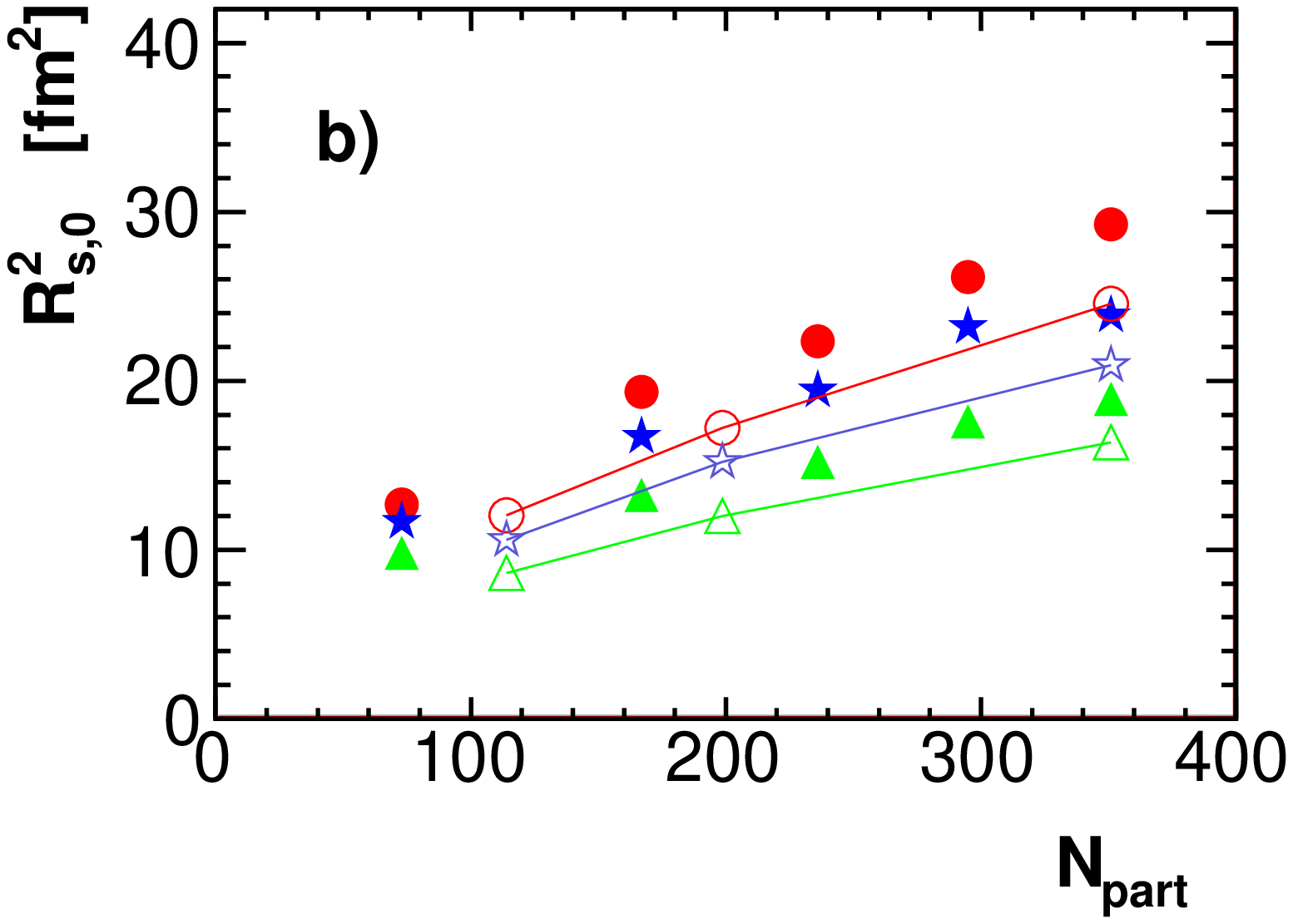} 
\includegraphics[angle=0,width=0.32 \textwidth]{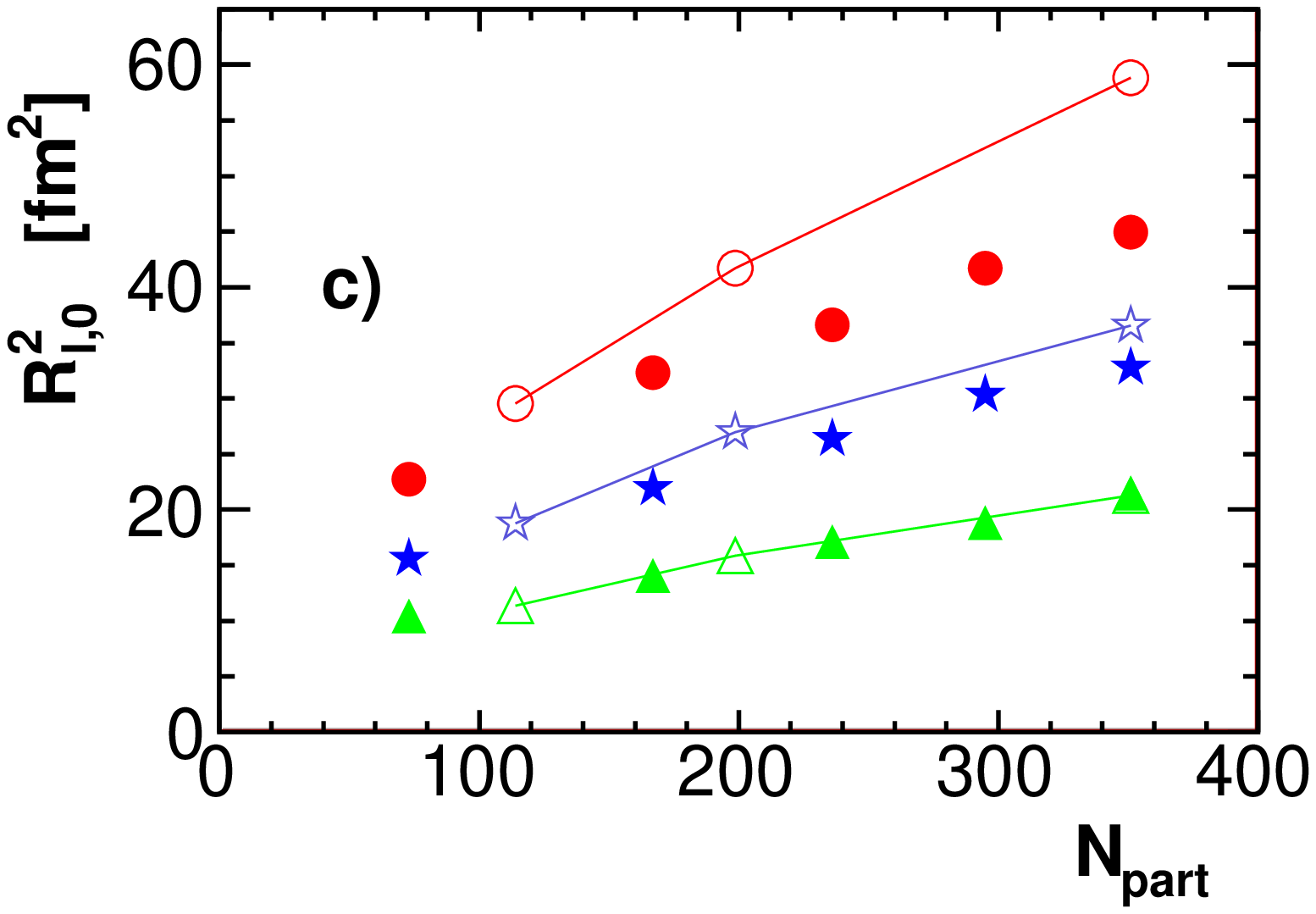} 
\caption{(Color online)   HBT radii (zeroth-order Fourier coefficients)
 with respect to second-order event plane for Au-Au collisions at $200$~GeV for different centralities,  $R_{o,0}^2$ (panel a), $R_{s,0}^2$ (panel b),
 and $R_{l,0}^2$ (panel d).
STAR collaboration data \cite{Adams:2003ra} are denoted by full symbols, circles  for $0.15$GeV$<k_\perp<0.25$GeV, 
stars  for $0.25$GeV$<k_\perp<0.35$GeV, and  triangles  for $0.35$GeV$<k_\perp<0.65$GeV.
Results of event-by-event
 hydrodynamic calculations are  shown with open symbols connected with lines.
\label{fig:au0}} 
\end{figure}

\begin{figure}
\includegraphics[angle=0,width=0.32 \textwidth]{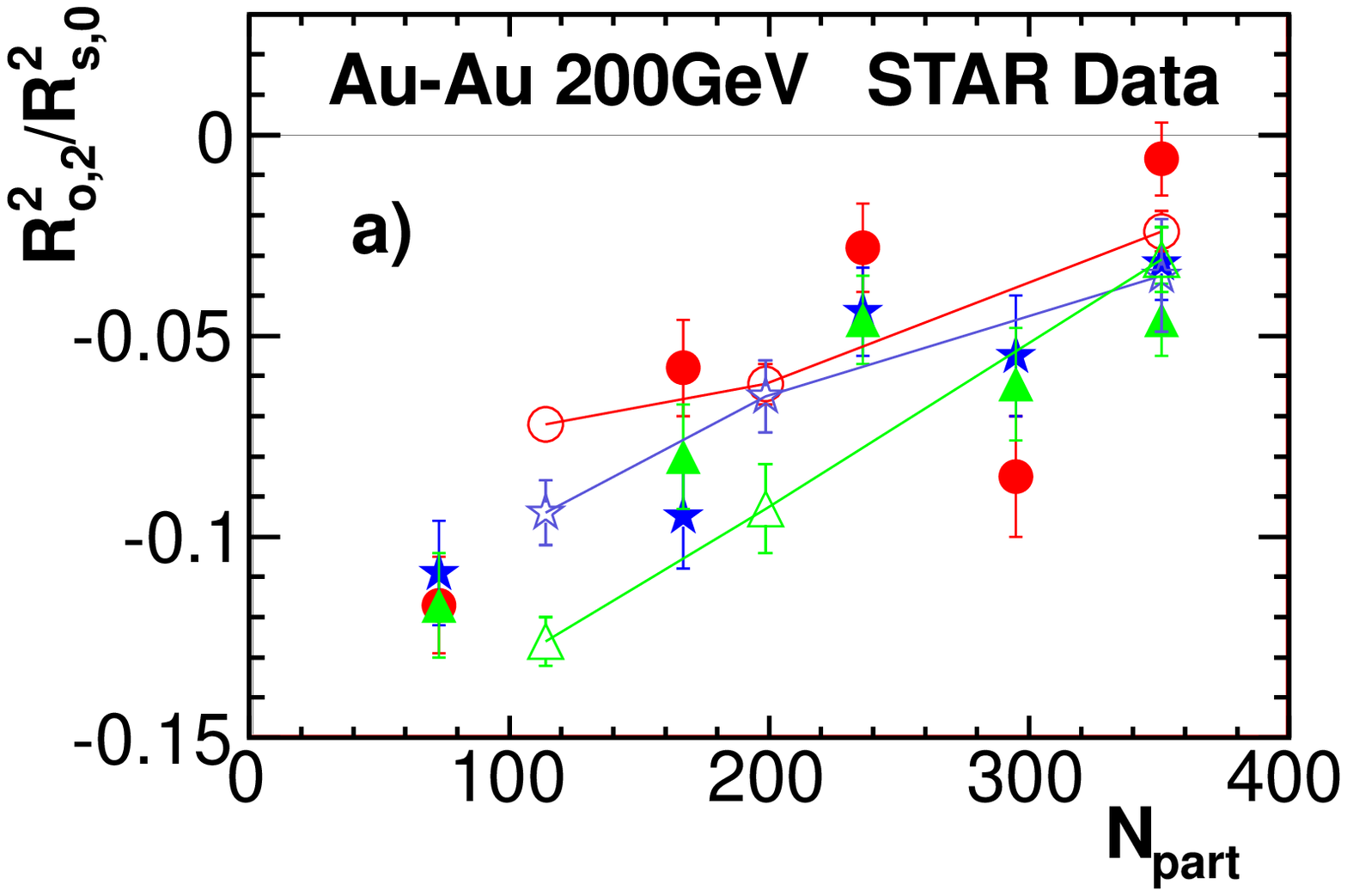} 
\includegraphics[angle=0,width=0.32 \textwidth]{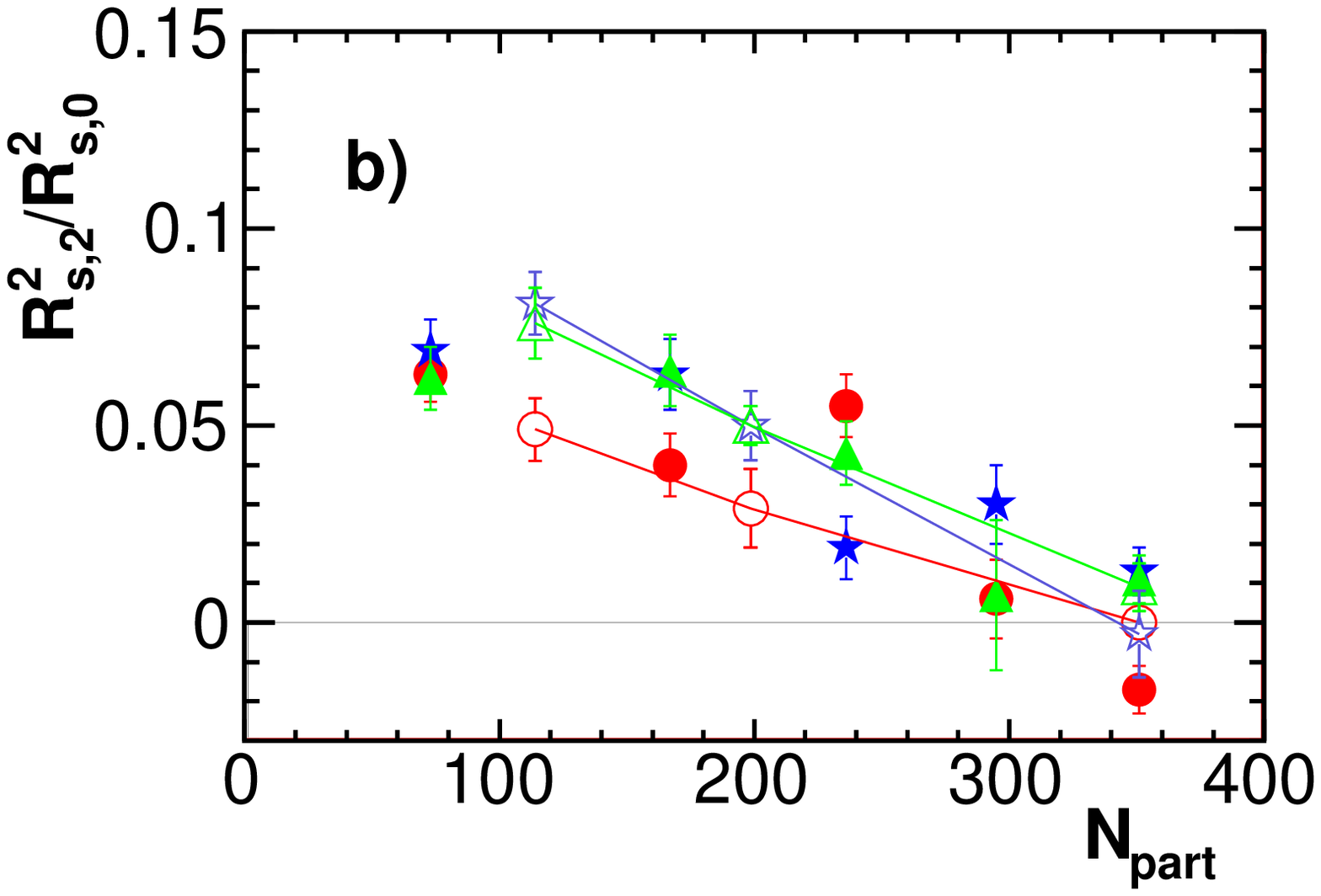} 
\includegraphics[angle=0,width=0.32 \textwidth]{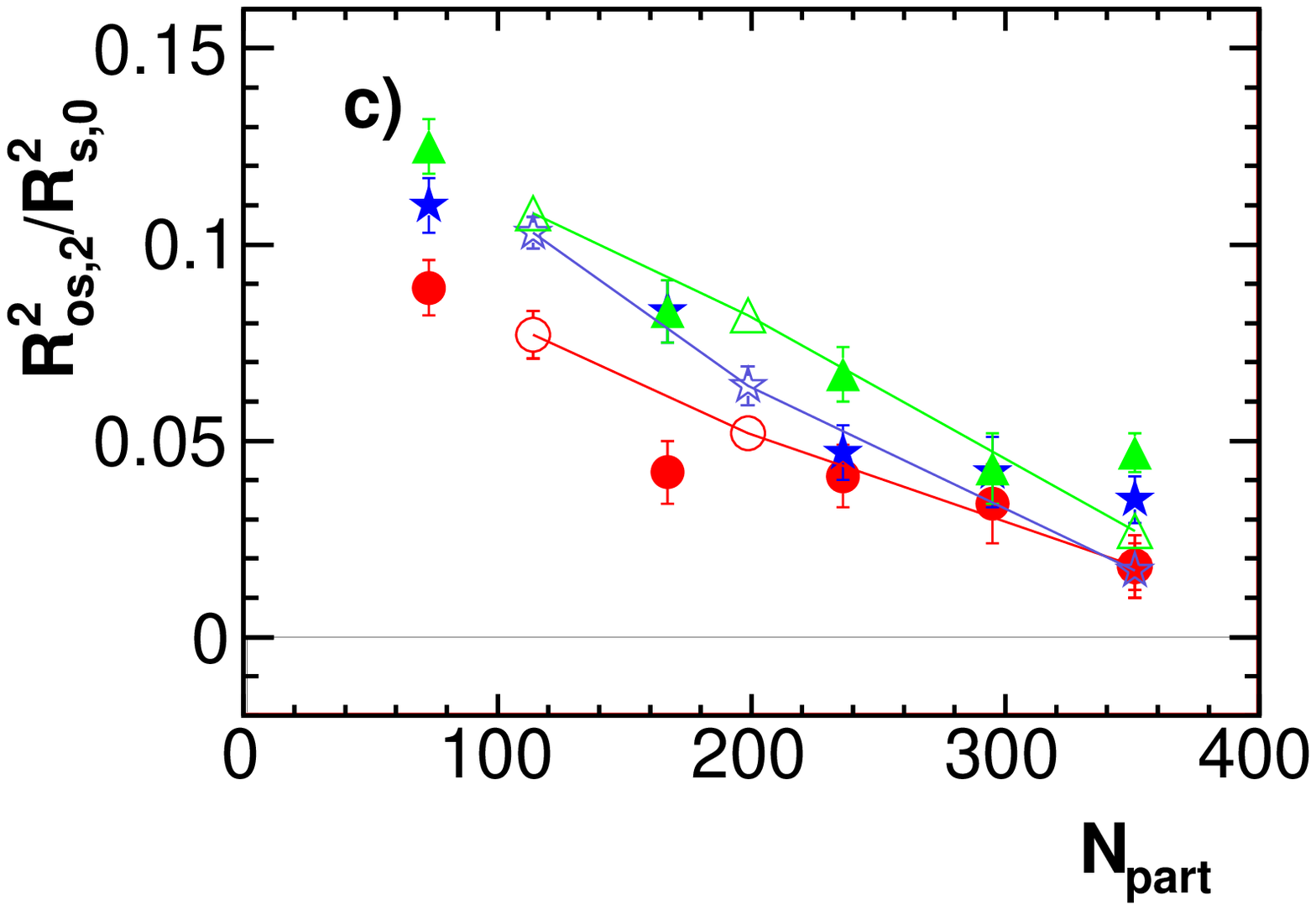} 
\caption{(Color online)  Second-order Fourier coefficients of the oscillations of the  HBT radii 
 with respect to the second-order event plane for Au-Au collisions at $200$~GeV for different centralities,  
$R_{o,2}^2/R_{s,0}^2$ (panel a),  $R_{s,2}^2/R_{s,0}^2$ (panel b),   $R_{os,2}^2/R_{s,0}^2$ (panel c). 
STAR collaboration data \cite{Adams:2003ra} are compared to results of hydrodynamic calculations, same symbols as
 in Fig. \ref{fig:au0}.
\label{fig:au2}} 
\end{figure}

In the experimental analysis, the correlation function is constructed using pairs from the
 same event in the numerator of Eq. \ref{eq:cq} and  pairs from mixed events
 for the denominator. In simulations,
 due to lower statistics, we combine $N_e$ ($100$ to $1600$) events corresponding  the same  
freeze-out hypersurface \cite{Bozek:2012hy}, where each freeze-out
surface is generated in a hydrodynamic evolution for a given initial condition. 
The numerator in correlation function (Eq. \ref{eq:cq}) is then  averaged over $N_h$ hypersurfaces.
 This corresponds to the averaging denoted by $\langle \dots \rangle$.
The  pairs from mixed events in the denominator are constructed from pions emitted from different freeze-out hypersurfaces. 
The histogram of the correlation function $C(q,k)$ for a given bin $q_a$, $k_b$ is
\begin{eqnarray}
C(q_a,k_b) =  &&\nonumber \\
\frac{\frac{1}{N_{pairs,num}}\sum_{j=1}^{N_h} \sum_{m\neq l=1  }^{N_e}\sum_{s=1}^{M_l}\sum_{f=1}^{M_m} \delta_{q_a}\delta_{k_b} \Psi(q,x_1-x_2)}
{\frac{1}{N_{pairs,den}}\sum_{i\neq j=1}^{N_h} \sum_{l,m=1  }^{N_e}\sum_{s=1}^{M_l}\sum_{f=1}^{M_m} \delta_{q_a}\delta_{k_b}} &&  \nonumber \\
&& , \label{eq:bincq}
\end{eqnarray}
where $M_l$ and $M_m$ are the pion multiplicities of the events $m$ and $l$ generated from the freeze-out
surface  $j$ (or $i$ and $j$ in the denominator), 
the symbols $\delta_{q_a}$ and $\delta_{k_b}$ are $1$ if the momenta 
 $q=p_s-p_f$ and $k=(p_s+p_f)/2$ fall into the respective bins and zero otherwise.

The direction of the second and third-order event plane $\psi$ is defined as the flow direction taken
 from $N_e$  combined THERMINATOR events  for each hydrodynamically generated freeze-out hypersurface $j$
\begin{equation}
v_n e^{i\psi_n^j}=\frac{\sum_{m=1}^{N_e}\sum_{s=1}^{M_m}e^{i n \phi_s}}{\sum_{m=1}^{N_e}\sum_{s=1}^{M_m} 1} \ ,
\end{equation}
where the sum is taken over  the charged particles with pseudorapidity 
$|\eta|<2$ ($\phi_s$ is the azimuthal direction of the particle transverse momentum),  $n=2,3$ for the second and third-order 
event plane respectively.
This procedure assures a very good event 
plane resolution. The azimuthal angles in the events generated from the hypersurface $j$
 are calculated relative to the event plane angle $\psi_n^j$.

\begin{figure}
\includegraphics[angle=0,width=0.32 \textwidth]{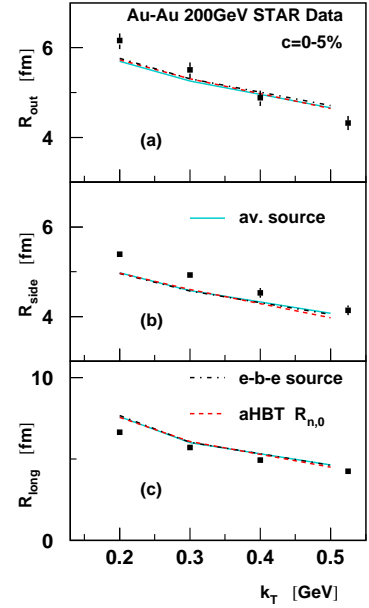} 
\caption{(Color online) 
The HBT radii $R_{o}$ (panel a), $R_{s}$ (panel b), and $R_{l}$ (panel c) for
Au-Au collisions at $200$~GeV, the squares denote the STAR collaboration data \cite{Adams:2004yc}. The hydrodynamic model results for event-by-event fluctuating emission
 sources (Eq. \ref{eq:cq}), for the averaged emission source (Eq. \ref{eq:cqav}), and for the zeroth-order radii from the azimuthally sensitive HBT analysis are represented by the dashed-dotted, solid, and dashed lines respectively.
\label{fig:hbt}}
\end{figure}

The correlation functions $C(q)$
 are calculated in bins in the azimuthal angle $\Phi$ and transverse momentum $k_\perp$.
The relative momentum is decomposed into  $q_{long}$, $q_{out}$, and $q_{side}$, 
and the correlations function is fitted using the Bertsch-Pratt formula \cite{Bertsch:1989vn,Pratt:1986cc}. For 
azimuthally sensitive interferometry at central rapidity and for symmetric collisions we use 
\cite{Wiedemann:1997cr,Lisa:2000ip}
\begin{eqnarray}
&& C(q_{long}, q_{out}, q_{side})= \nonumber \\
&& 1+\lambda e^{-R_o^2 q_{out}^2-R_s^2q_{side}^2-R_l^2q_{long}^2-2R_{os}^2 q_{out}q_{side}} 
\ . \label{eq:berpra}
\end{eqnarray}
The centrality and $k_\perp$ bins are similar as used
 in the STAR, PHENIX and ALICE experiments, we
use $6$ bins in the azimuthal direction between $0$ and  $\pi$ for the  second-order event plane 
analysis and between $0$ and $2\pi/3$ for the third-order event plane analysis.
The angular dependence is fitted using the zeroth and second- (third-)
 order harmonics for the $n=2$ ($n=3$)
event plane  analysis
\begin{eqnarray}
 R_o^2(\Phi)&=&R_{o,0}^2+2R_{o,n}^2\cos(n\Phi) \nonumber \\
 R_s^2(\Phi)&=&R_{s,0}^2+2R_{s,n}^2\cos(n\Phi) \nonumber \\
 R_{os}^2(\Phi)&=&2R_{os,n}^2\sin(n\Phi) \ . \label{eq:rphi}
\end{eqnarray}
In all cases, the obtained angular dependence of $R_l(\Phi)$ is consistent with a constant value $R_{l,0}$.

\begin{figure}
\includegraphics[angle=0,width=0.32 \textwidth]{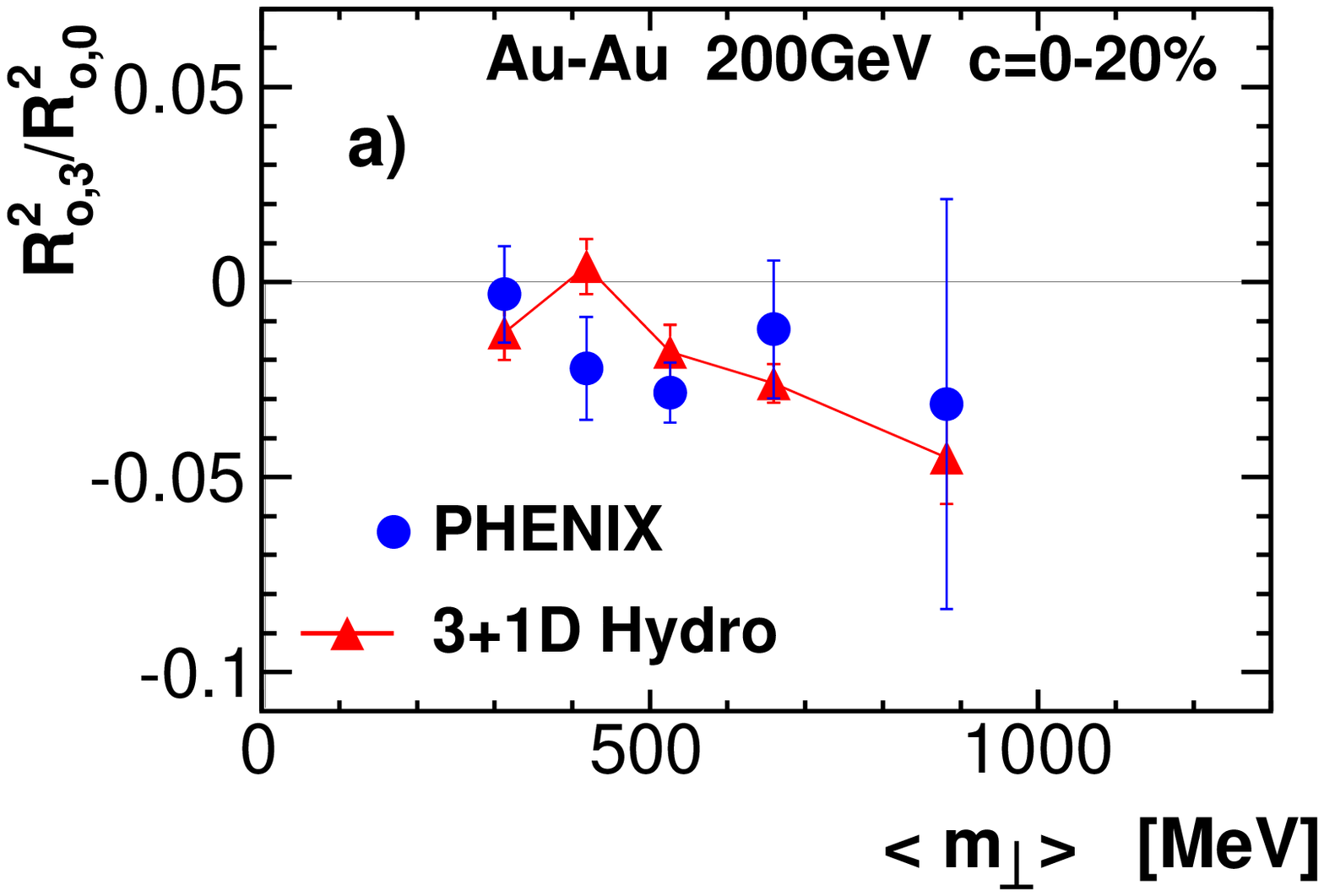} 
\includegraphics[angle=0,width=0.32 \textwidth]{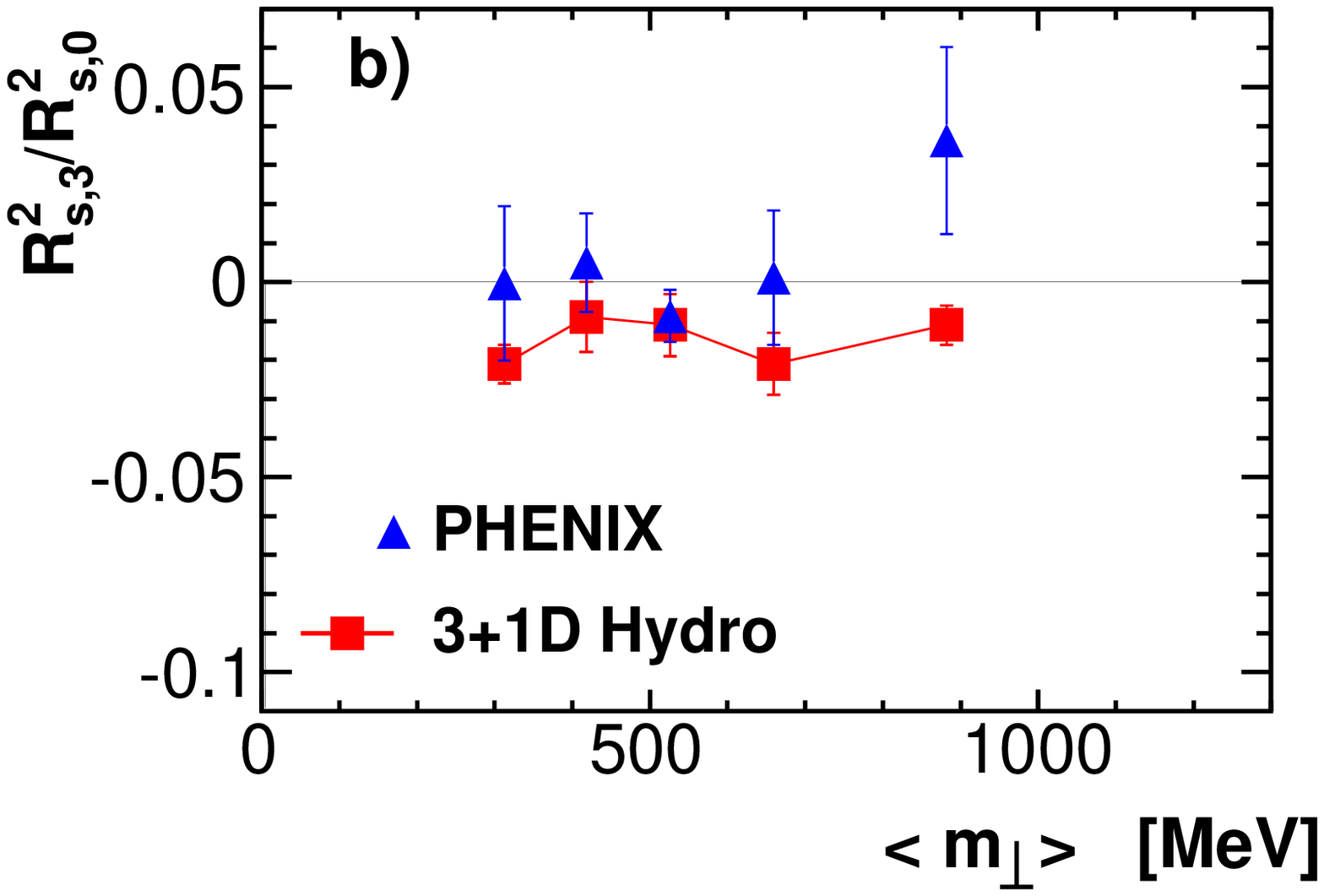} 
\caption{(Color online)   The third-order Fourier coefficients 
of the oscillations of the  HBT radii
 with respect to third-order event plane for Au-Au collisions at $200$~GeV for centrality $0$-$20$\% as function of the transverse mass. (Panel a)
$R_{o,3}^2/R_{o,0}^2$, preliminary PHENIX data \cite{niidawpcf} (circles) are compared to hydrodynamic calculations (triangles connected with lines). (Panel b)   $R_{s,3}^2/R_{s,0}^2$, preliminary  PHENIX data (triangles) are compared to hydrodynamic calculations (squares
 connected with lines).
\label{fig:c0au3}} 
\end{figure}

\begin{figure}
\includegraphics[angle=0,width=0.32 \textwidth]{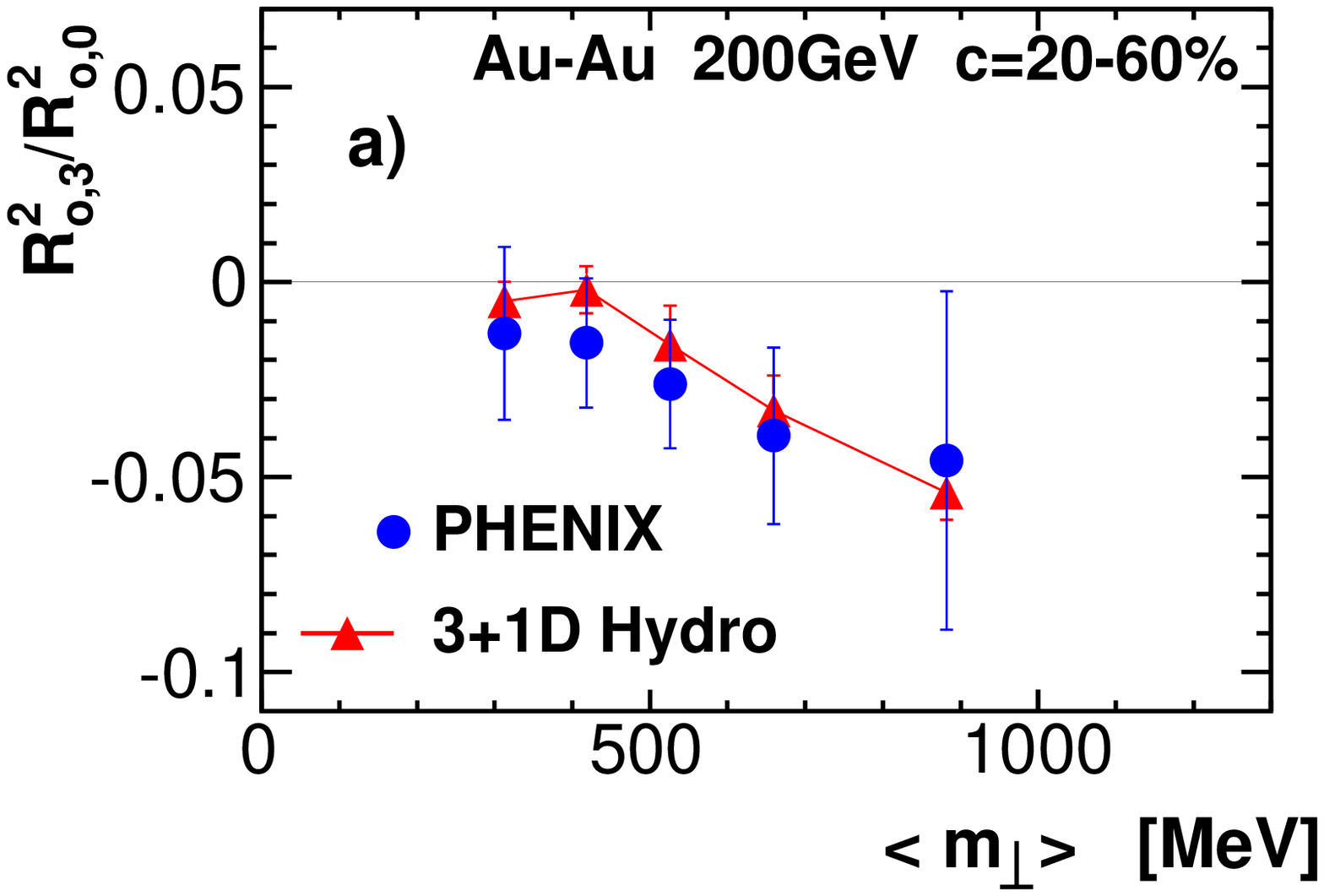} 
\includegraphics[angle=0,width=0.32 \textwidth]{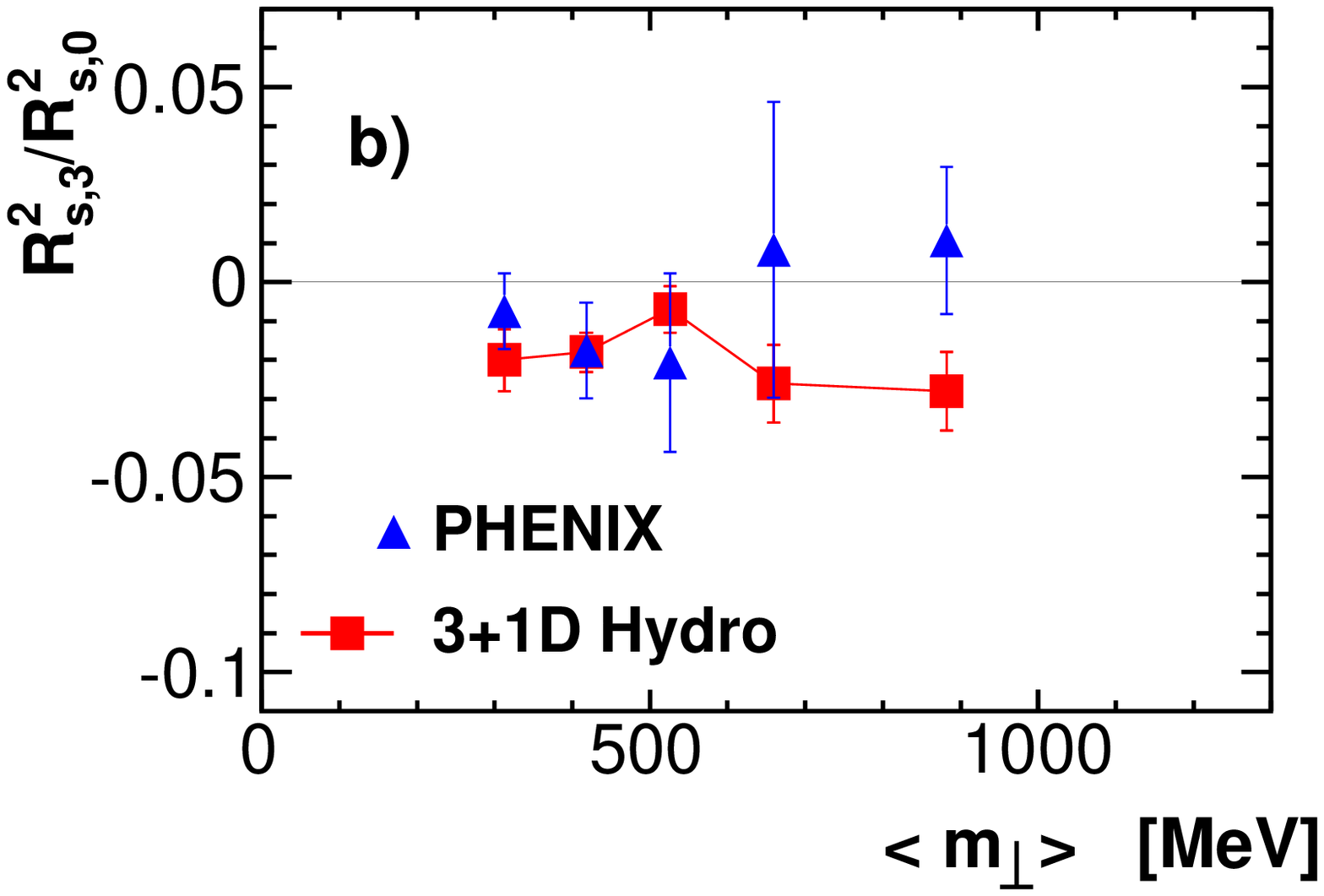} 
\caption{(Color online) Same as Fig. \ref{fig:c0au3} but for
 centrality $20$-$60$\%.
\label{fig:c20au3}} 
\end{figure}  

We begin with the analysis of the azimuthally sensitive interferometry 
with respect to the third-order event plane for the fireball with smooth 
initial conditions described in Sect. \ref{sec:model} (Eq. \ref{eq:ini3}).
In that case, there is one hydrodynamic freeze-out surface and  Eq.
 \ref{eq:bincq} is modified, the  sum over the freeze-out hypersurfaces 
is absent.
The scaled third-order HBT radii $R_{i,3}^2$ are shown in Fig. \ref{fig:model}.
The azimuthal modulation is the largest for the {\it out} radius $R_{o,3}^2/R_{o,0}^2$.
It is negative and its magnitude increases with the transverse momentum. 
The {\it side} radius $R_{s,3}^2$ is close to zero and negative. $R_{os,3}^2$ is positive and increases with $k_\perp$. The shape of the 
geometry of the freeze-out hypersurface is close to azimuthal symmetry
 for most of the 
evolution (Fig. \ref{fig:cont}). The third-order HBT radii are close to the 
deformed flow scenario \cite{Plumberg:2013nga}. We notice that  the scaled 
third-order 
HBT radii do not change substantially when  the freeze-out temperature is 
lowered from
 $150$ to $140$~MeV.

\begin{figure}
\includegraphics[angle=0,width=0.32 \textwidth]{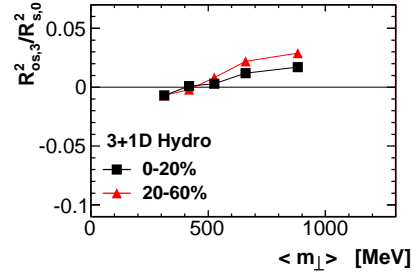} 
\caption{(Color online) (Color online)  The third-order Fourier coefficients 
of the oscillations of the  HBT  radius $R_{os,3}^2/R_{s,0}^2$
 with respect to third-order event plane for Au-Au collisions at $200$~GeV.
 Hydrodynamic simulation results for centrality $0$-$20$\% and $20$-$60$\% are denoted with squares and triangles respectively.
\label{fig:c020au3}} 
\end{figure}

\begin{figure}
\includegraphics[angle=0,width=0.32 \textwidth]{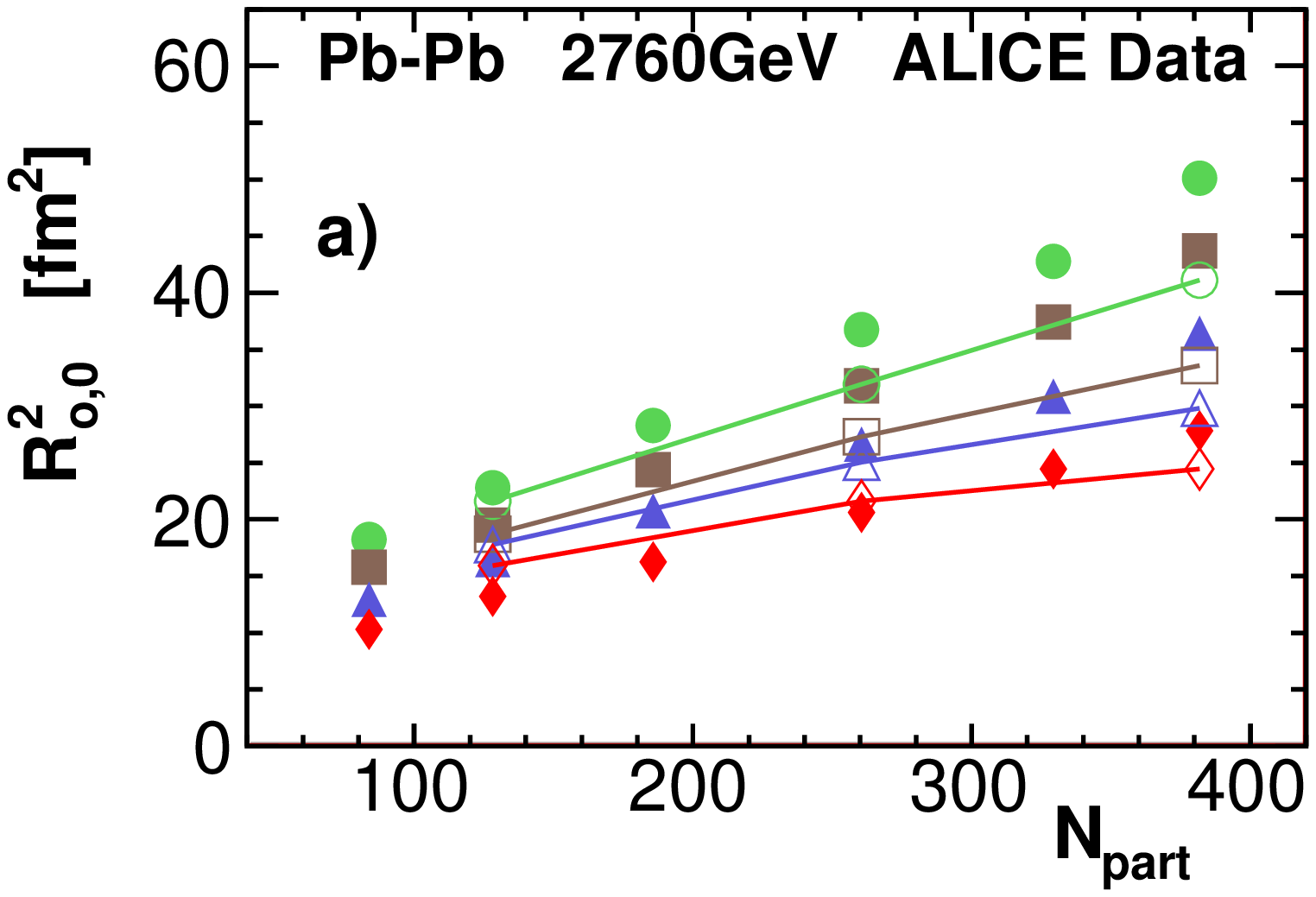} 
\includegraphics[angle=0,width=0.32 \textwidth]{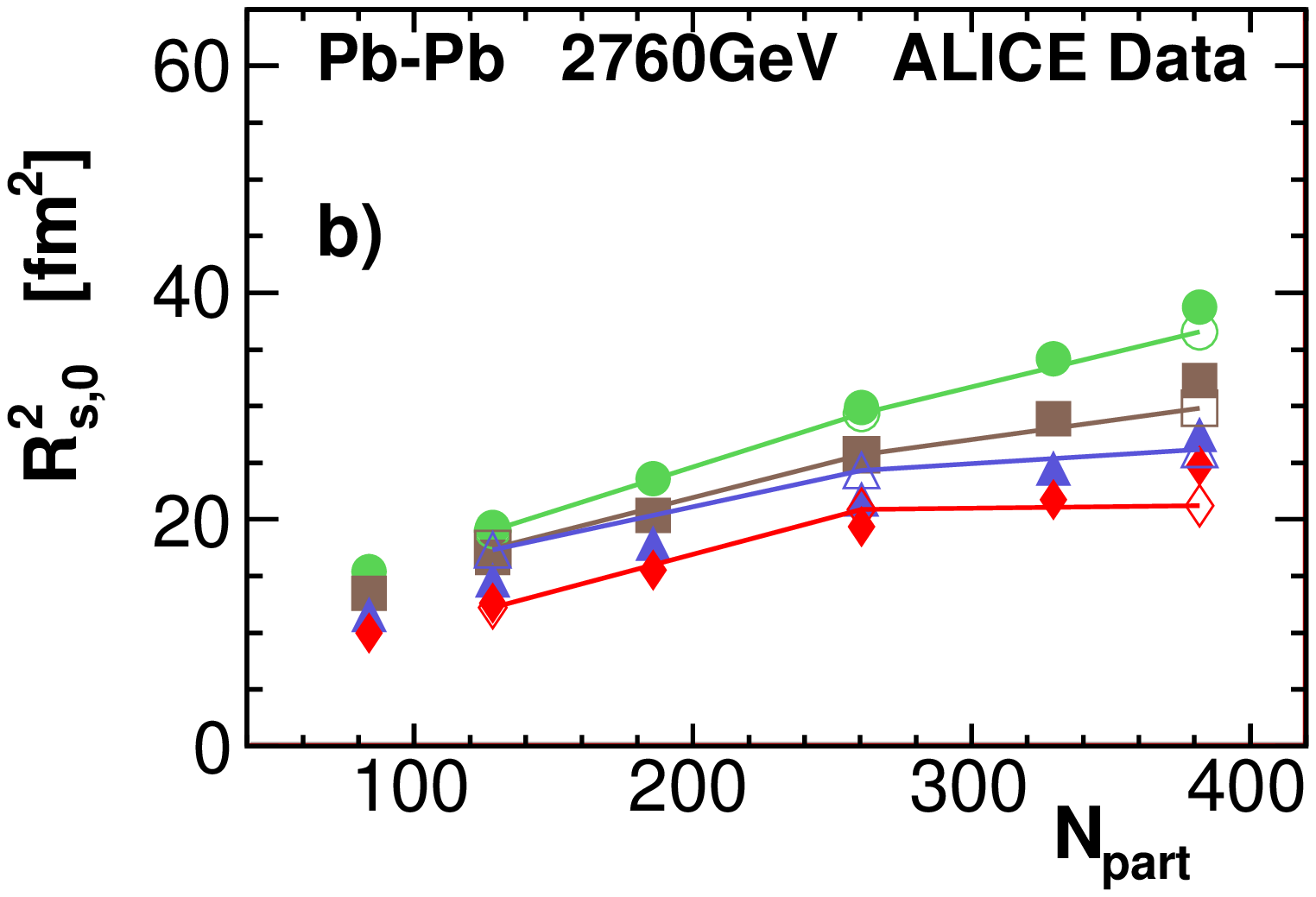} 
\includegraphics[angle=0,width=0.32 \textwidth]{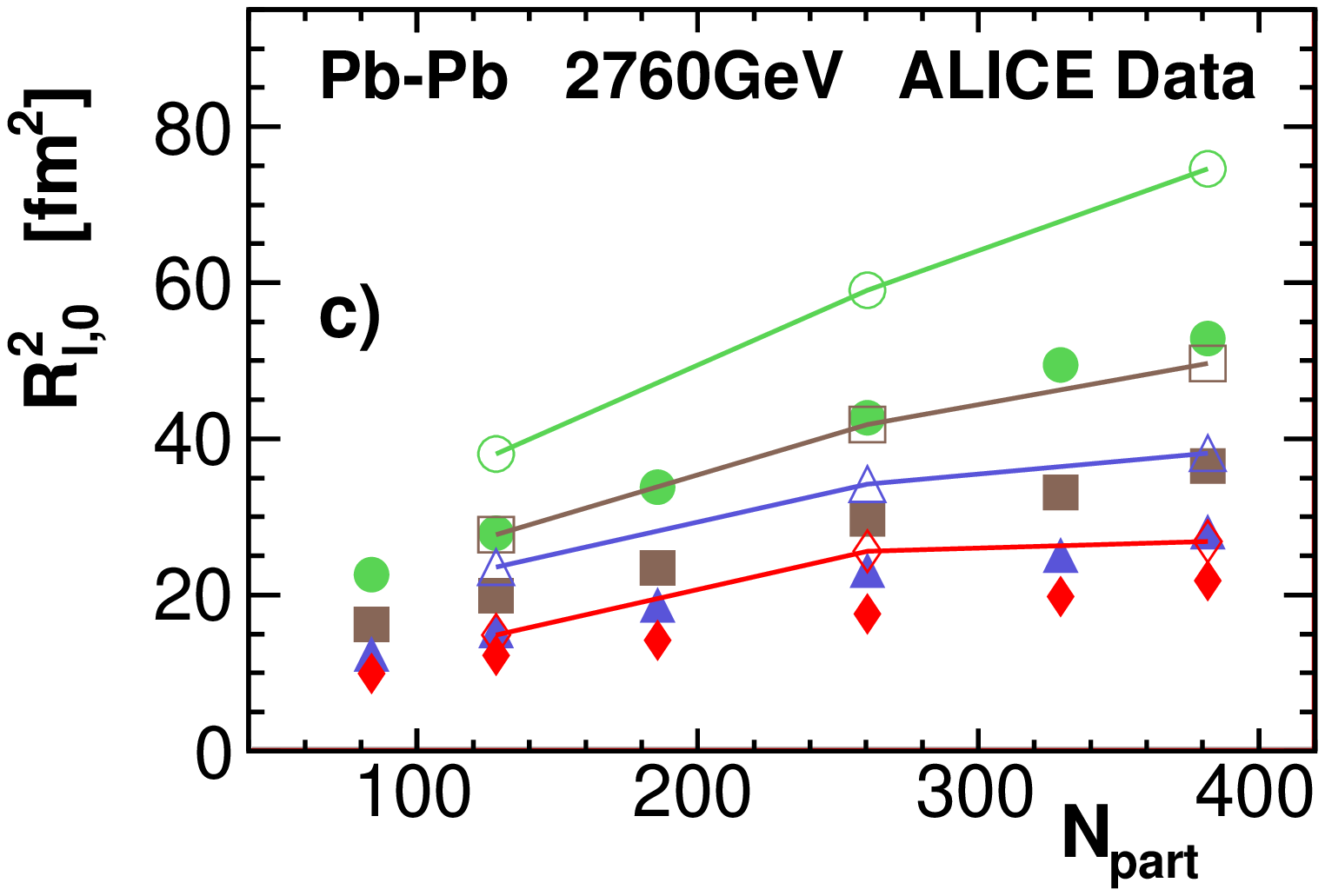} 
\caption{(Color online) HBT radii  (zeroth-order Fourier coefficients)
 with respect to the second-order event plane for Pb-Pb collisions at $2.76$~TeV for different centralities,  $R_{o,0}^2$ (panel a), $R_{s,0}^2$ (panel b),
 and $R_{l,0}^2$ (panel c). Preliminary
ALICE collaboration data \cite{logginswpcf} are denoted by full symbols, circles  for $0.2$GeV$<k_\perp<0.3$GeV, squares  for $0.3$GeV$<k_\perp<0.4$GeV, 
triangles  for $0.4$GeV$<k_\perp<0.5$GeV, and  diamonds  for $0.5$GeV$<k_\perp<0.7$GeV.
Results of event-by-event
 hydrodynamic calculations are  shown with open symbols connected with lines.
\label{fig:pb0}} 
\end{figure}

\begin{figure}
\includegraphics[angle=0,width=0.32 \textwidth]{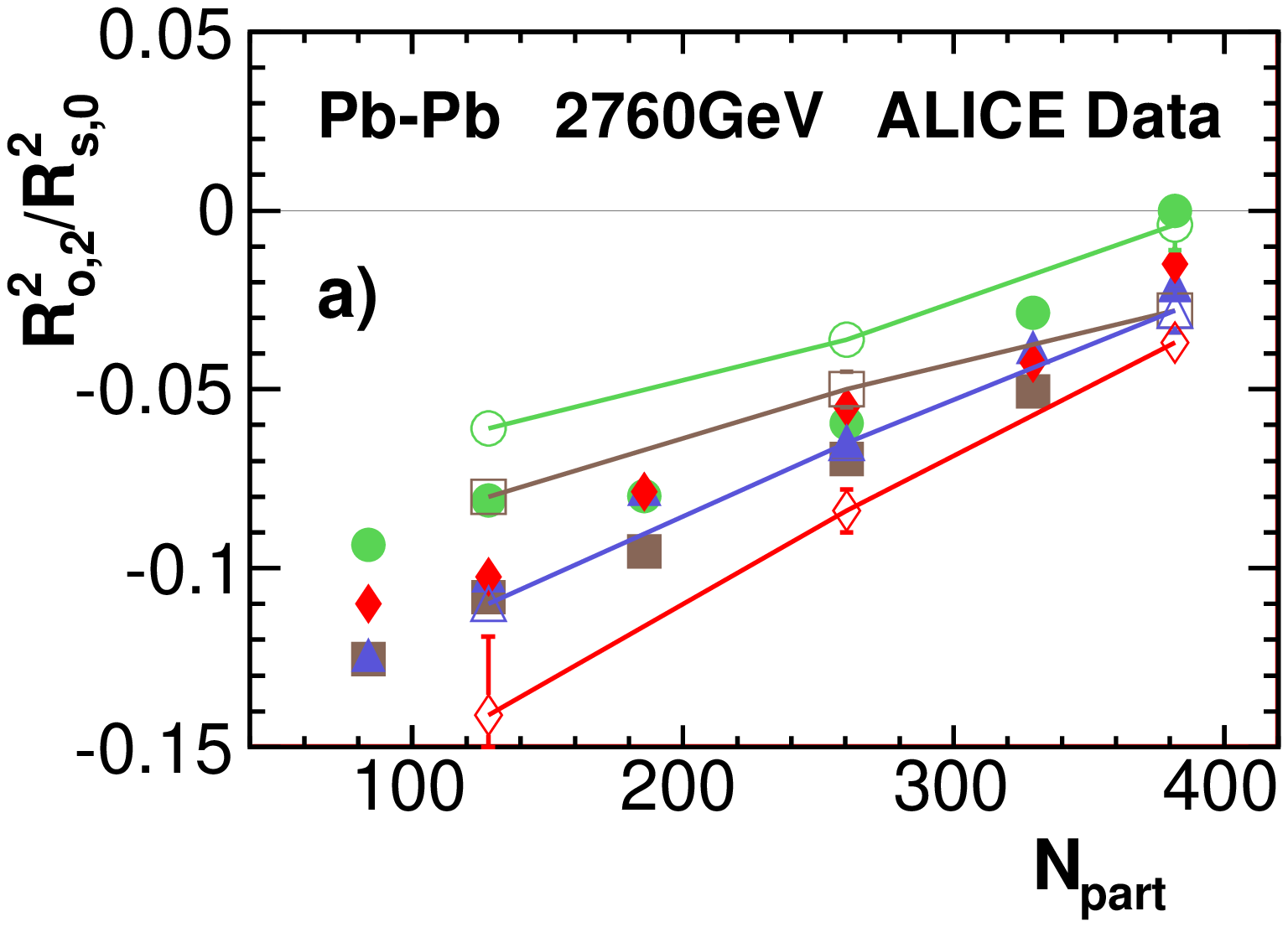} 
\includegraphics[angle=0,width=0.32 \textwidth]{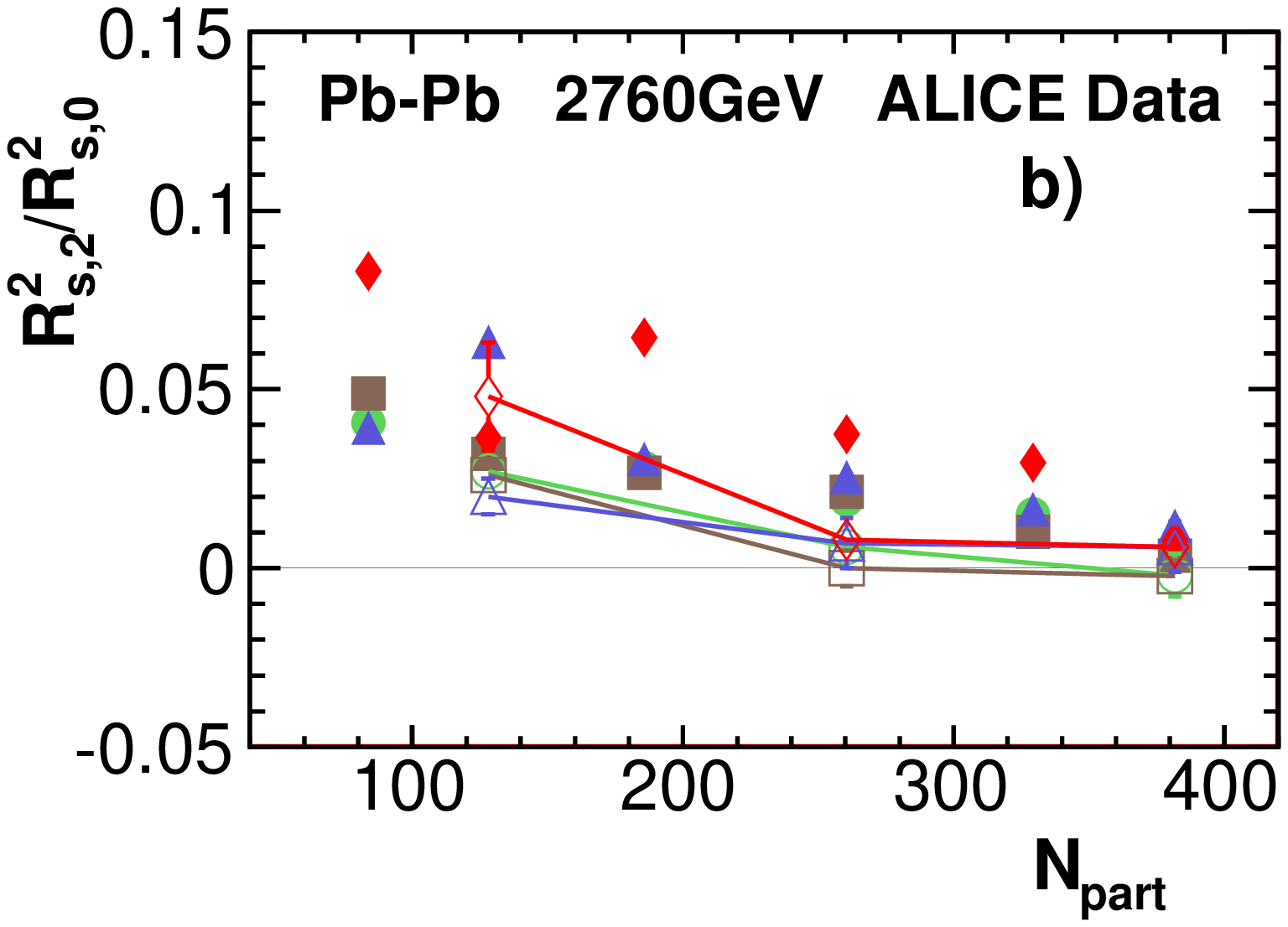} 
\includegraphics[angle=0,width=0.32 \textwidth]{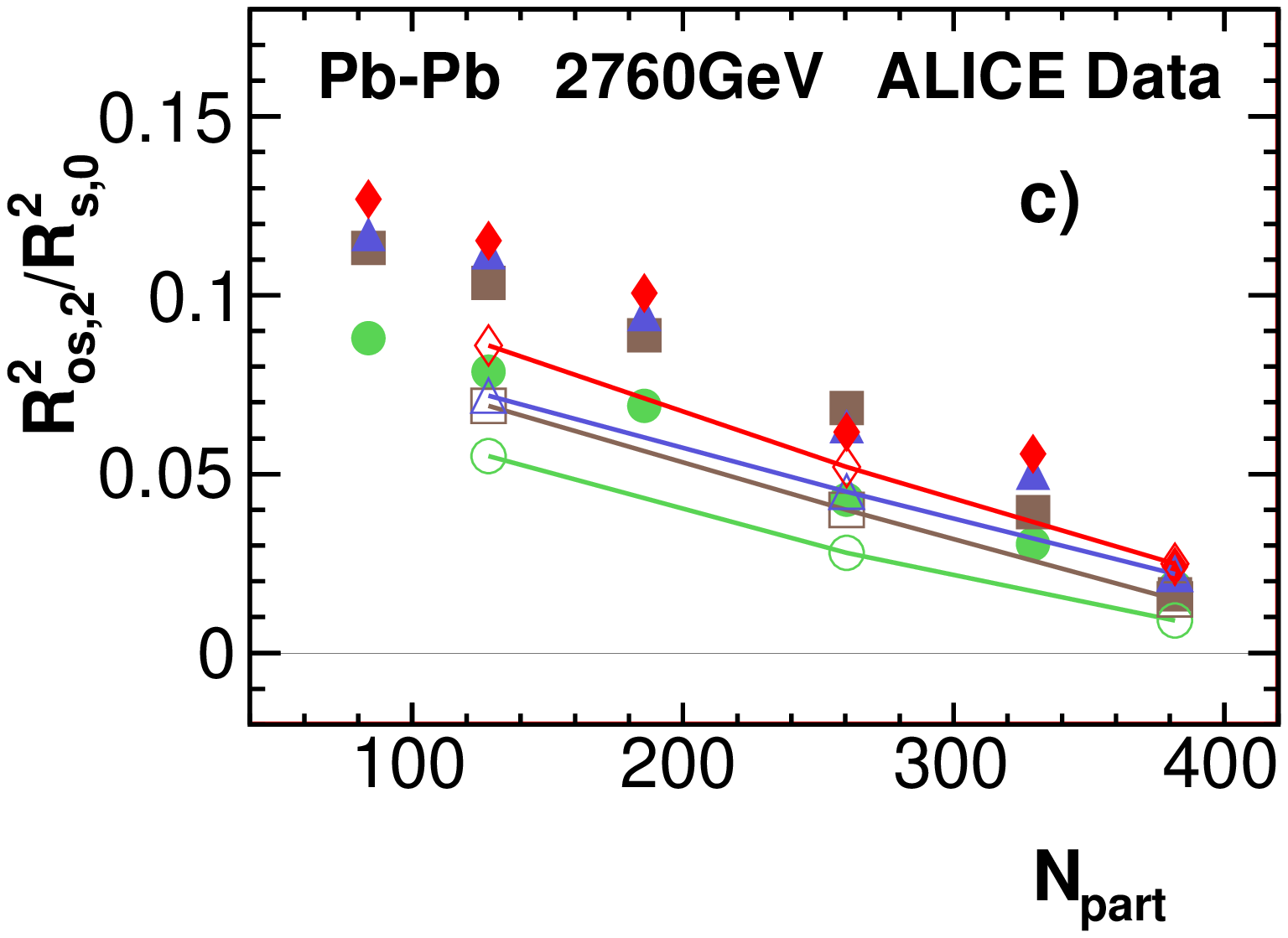} 
\caption{(Color online)  The second-order Fourier coefficients of the 
oscillations of the  HBT radii 
 with respect to second-order event plane for Pb-Pb collisions at $2.76$~TeV for different centralities,  
$R_{o,2}^2/R_{s,0}^2$ (panel a),  $R_{s,2}^2/R_{s,0}^2$ (panel b), and  $R_{os,2}^2/R_{s,0}^2$ (panel c). Preliminary
ALICE collaboration data \cite{logginswpcf} are compared to results of hydrodynamic calculations, same symbols as
 in Fig. \ref{fig:pb0}.
\label{fig:pb2}} 
\end{figure}   

\begin{figure}
\includegraphics[angle=0,width=0.32 \textwidth]{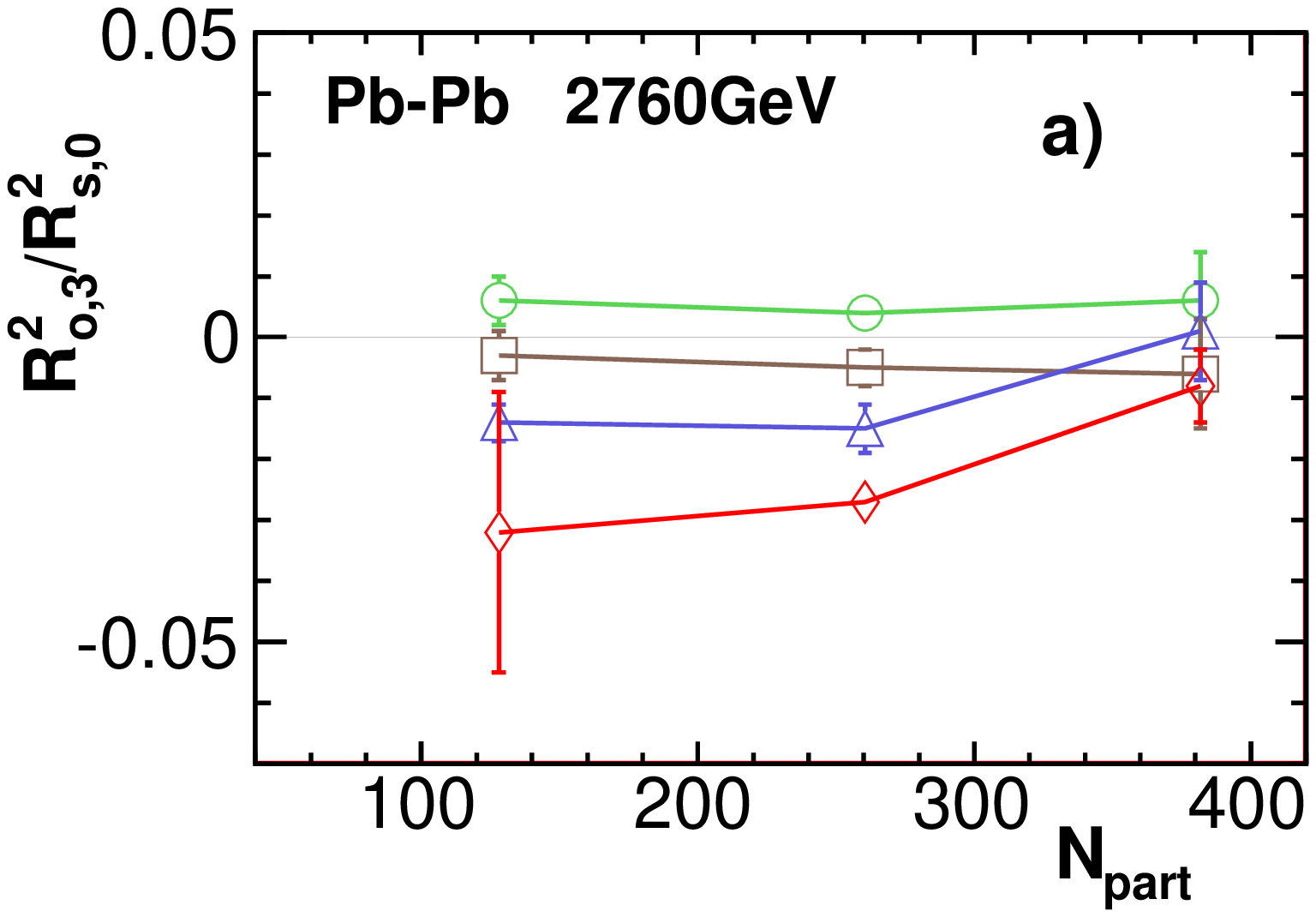} 
\includegraphics[angle=0,width=0.32 \textwidth]{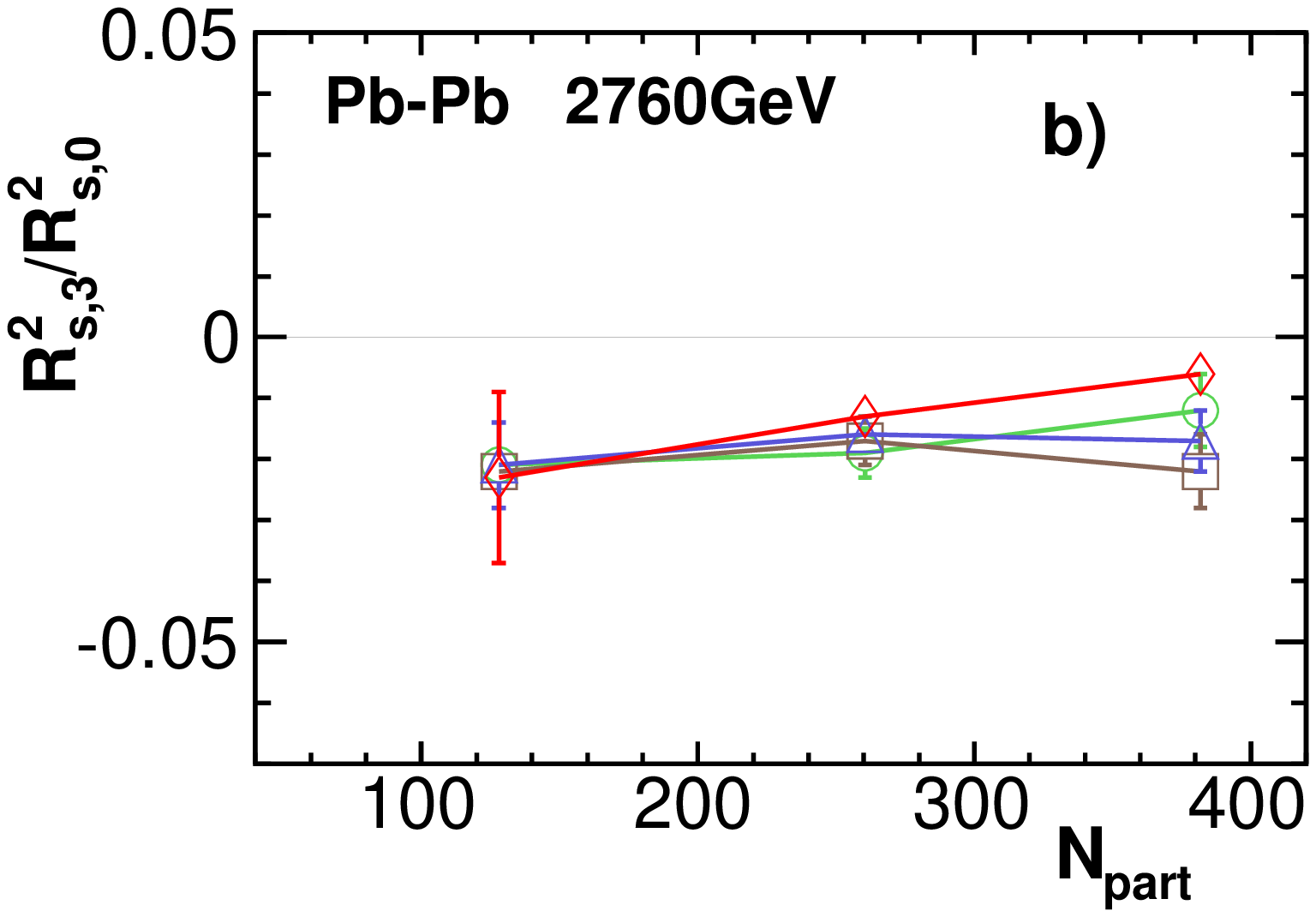} 
\includegraphics[angle=0,width=0.32 \textwidth]{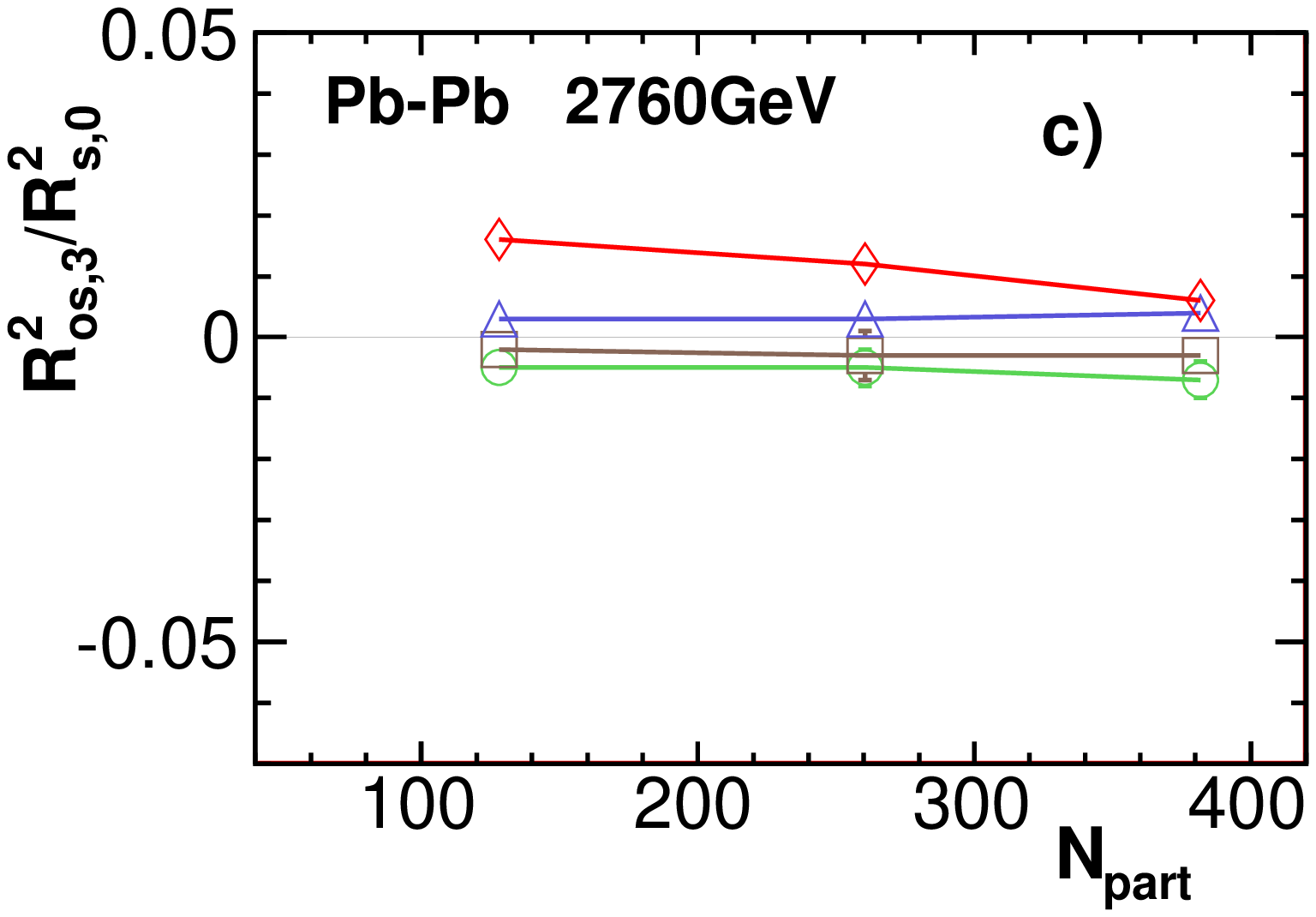} 
\caption{(Color online)  The third-order Fourier coefficients of the oscillations of the  HBT radii 
 with respect to third-order event plane for Pb-Pb collisions at $2.76$~TeV for different centralities,  
$R_{o,3}^2/R_{s,0}^2$ (panel a),  $R_{s,3}^2/R_{s,0}^2$ (panel b),  and $R_{os,3}^2/R_{s,0}^2$ (panel c). 
 Results of hydrodynamic calculations are denoted with the  same symbols as
 in Fig. \ref{fig:pb0}.
\label{fig:pb3}} 
\end{figure}   

The second-order azimuthally sensitive HBT radii for Au-Au collisions at $200$~GeV
have been analyzed by the STAR
 collaboration \cite{Adams:2003ra}. The ideal fluid 
hydrodynamic
 calculation, using smooth initial conditions \cite{Kisiel:2008ws} reproduce
 semiquantitatively 
the data. In Figs. \ref{fig:au0} and \ref{fig:au2} are presented the 
results of  simulations using the  event-by-event $3+1$-D 
viscous hydrodynamic model. 
The zeroth-order  radii are in semiquantitative agreement with the experimental 
measurements (Fig. \ref{fig:au0}). For central collisions  ($0$-$5$\%) the angle 
averaged HBT radii are the same as the zeroth-order radii $R_{i,0}$  (Fig. \ref{fig:hbt}).
The deviations from the data are $10$-$15$\%, similar as in other 
calculations of the angle averaged HBT radii \cite{Broniowski:2008vp}. In Fig. 
\ref{fig:hbt} are shown also the results of a calculation using pairs of pions
 from different hypersurfaces in the numerator of Eq. \ref{eq:bincq}. This corresponds
to the use of the event averaged emission function in the numerator of Eq. \ref{eq:cq}
\begin{eqnarray}
 && C_{av}(q,k)= \nonumber \\
&& \frac{ \int d^4x_1 d^4x_2 \langle S(x_1,p_1)\rangle \langle S(x_2,p_2)\rangle 
|\Psi(k,(x_1-x_2))|^2}
{\int   d^4x_1 \langle S(x_1,p_1)\rangle \int  d^4x_2 \langle S(x_2,p_2) \rangle} \ . \nonumber \\  
\label{eq:cqav}
\end{eqnarray}
The extracted radii are within $3$\% from the ones obtained using  the correct 
 event-by-event 
average as in  Eq. \ref{eq:cq}. The second harmonics of the HBT radii  are in fair agreement with the experimental results (Fig. \ref{fig:au2}), both in sign and in magnitude.
The second-order components of the radii increase with centrality, indicating 
a stronger deformation of the flow geometry in peripheral events.

The azimuthally sensitive HBT radii with respect to the third-order event plane are calculated for centralities $0$-$20$\% and $20$-$60$\% and compared to preliminary results of the PHENIX collaboration \cite{logginswpcf}
 (Figs. \ref{fig:c0au3} and \ref{fig:c20au3}). The scaled radius $R_{o,3}^2/R_{o,0}^2$ is 
negative and  similar as in the experimental data. The agreement is slightly better for the 
semiperipheral events  (Fig. \ref{fig:c20au3}, panel a).
 Its magnitude increases with the transverse mass. The ratio 
$R_{s,3}^2/R_{s,0}^2$ 
has a small, negative value. The experimental
 data points are spread around zero, without a clear 
systematics of  the sign. The third-order harmonic of the $R_{os,3}^2$ radius is positive,
 increasing with the transverse mass, and larger for the more peripheral events
(Fig. \ref{fig:c020au3}). We observe that the results of the realistic, 
event-by-event  calculation  for the azimuthally sensitive HBT radii are similar as for the model calculation with smooth initial conditions presented above (Fig. \ref{fig:model}). There is no significant centrality dependence of the third-harmonic 
component of the HBT radii, unlike for the HBT radii with 
respect to the second-order event plane. This reflects a weaker dependence 
of the triangular flow on centrality as compared to the elliptic flow.

The azimuthally sensitive HBT radii with respect to the second-order event plane are calculated for Pb-Pb collisions at $2.76$~TeV. The zeroth and second-order harmonics are plotted in Figs. \ref{fig:pb0} and \ref{fig:pb2}. We notice that the calculations are in 
semiquantitative agreement with the preliminary ALICE collaboration 
data \cite{logginswpcf}. The sign of the measured and calculated second 
harmonics  of the HBT
 radii are similar as for Au-Au collisions at $200$~GeV (Fig. \ref{fig:au2}). 
The centrality dependence of the second harmonic component of the HBT radii
is strong, as expected from the strong centrality dependence of the elliptic flow.

The predictions of the azimuthally sensitive HBT interferometry 
analysis with respect to the third-order event plane for Pb-Pb collisions
are presented in Fig. \ref{fig:pb3}.
The results are similar as for  Au-Au collisions at RHIC energies. The radii
$R_{o,3}^2$ and $R_{s,3}^2$ are negative. The radii $R_{o,3}^2/R_{s,0}^2$ 
have a larger magnitude,
increasing with the transverse momentum and centrality, while $R_{s,3}^2/R_{s,0}^2$
has no noticeable $k_\perp$ dependence. $R_{os,3}^2/R_{s,0}^2$ is positive,
 increasing with the transverse momentum.

\section{Summary}

The azimuthally sensitive 
 interferometry analysis is performed in the event-by-event
$3+1$-D viscous hydrodynamic model. We construct the two-pion correlation
 function and calculate the azimuthal dependence of the HBT radii. 
Two cases are studied, the   radii defined with respect to the 
second and  third-order event planes. For the first case
 the azimuthal dependence 
of the radii is decomposed into a zeroth and second harmonic, while in
 the second case we use the zeroth and third harmonic. 

The HBT radii calculated with respect to the second-order event plane are compared to the experimental results of the STAR and ALICE collaborations for Au-Au
collisions at $200$~GeV \cite{Adams:2003ra} and Pb-Pb collisions 
at $2.76$~TeV \cite{logginswpcf}. The sign and the magnitude of the calculated
 second-order harmonics of
 the HBT radii is in semiquantitative agreement with the experimental results. 
We find a strong centrality dependence of the scaled ratios $R_{o,2}^2/R_{s,0}^2$,
 $R_{s,2}^2/R_{s,0}^2$, and $R_{os,2}^2/R_{s,0}^2$.

The third harmonic of the azimuthal dependence of the HBT radii 
with respect  to the third-order event plane is calculated for RHIC and 
LHC energies. AT RHIC energies, we 
find a negative value of $R_{o,3}^2/R_{o,0}^2$, with a magnitude
 increasing with the transverse momentum, and with 
weak centrality dependence. These effects are
in agreement with preliminary PHENIX collaboration data \cite{niidawpcf}.
The calculated  $R_{s,3}^2/R_{s,0}^2$ is small and negative, with 
no significant dependence on transverse momentum.
The predicted value of $R_{os,3}^2/R_{s,0}^2$ is positive and 
increases with the transverse momentum. The predictions for the 
third-order HBT radii in Pb-Pb collisions at $2.76$~TeV are qualitatively similar.

During the dynamic expansion of the fireball the initial triangular deformation
is washed-out, or even reversed. The model gives 
 large magnitude, negative values of $R_{o,3}^2$ and small magnitude,
 negative 
values of $R_{s,3}^2$. Detailed experimental and model studies may 
 serve as a way to establish the size and the sign  of the
 triangular deformation at freeze-out.

\begin{acknowledgments}
Supported  by National Science Centre, grant
DEC-2012/05/B/ST2/02528 and by PL-Grid Infrastructure.                                  
\end{acknowledgments}

\bibliography{../hydr}

\begin{thebibliography}{10}%
\makeatletter
\providecommand \@ifxundefined [1]{%
 \ifx #1\undefined \expandafter \@firstoftwo
 \else \expandafter \@secondoftwo
\fi
}%
\providecommand \@ifnum [1]{%
 \ifnum #1\expandafter \@firstoftwo
 \else \expandafter \@secondoftwo
\fi
}%
\providecommand \enquote [1]{``#1''}%
\providecommand \bibnamefont  [1]{#1}%
\providecommand \bibfnamefont [1]{#1}%
\providecommand \citenamefont [1]{#1}%
\providecommand\href[0]{\@sanitize\@href}%
\providecommand\@href[1]{\endgroup\@@startlink{#1}\endgroup\@@href}%
\providecommand\@@href[1]{#1\@@endlink}%
\providecommand \@sanitize [0]{\begingroup\catcode`\&12\catcode`\#12\relax}%
\@ifxundefined \pdfoutput {\@firstoftwo}{%
 \@ifnum{\z@=\pdfoutput}{\@firstoftwo}{\@secondoftwo}%
}{%
 \providecommand\@@startlink[1]{\leavevmode\special{html:<a href="#1">}}%
 \providecommand\@@endlink[0]{\special{html:</a>}}%
}{%
 \providecommand\@@startlink[1]{%
  \leavevmode
  \pdfstartlink
   attr{/Border[0 0 1 ]/H/I/C[0 1 1]}%
   user{/Subtype/Link/A<</Type/Action/S/URI/URI(#1)>>}%
  \relax
 }%
 \providecommand\@@endlink[0]{\pdfendlink}%
}%
\providecommand \url  [0]{\begingroup\@sanitize \@url }%
\providecommand \@url [1]{\endgroup\@href {#1}{\urlprefix}}%
\providecommand \urlprefix [0]{URL }%
\providecommand \Eprint[0]{\href }%
\@ifxundefined \urlstyle {%
  \providecommand \doi [1]{doi:\discretionary{}{}{}#1}%
}{%
  \providecommand \doi [0]{doi:\discretionary{}{}{}\begingroup
  \urlstyle{rm}\Url }%
}%
\providecommand \doibase [0]{http://dx.doi.org/}%
\providecommand \Doi[1]{\href{\doibase#1}}%
\providecommand \bibAnnote [3]{%
  \BibitemShut{#1}%
  \begin{quotation}\noindent
    \textsc{Key:}\ #2\\\textsc{Annotation:}\ #3%
  \end{quotation}%
}%
\providecommand \bibAnnoteFile [2]{%
  \IfFileExists{#2}{\bibAnnote {#1} {#2} {\input{#2}}}{}%
}%
\providecommand \typeout [0]{\immediate \write \m@ne }%
\providecommand \selectlanguage [0]{\@gobble}%
\providecommand \bibinfo [0]{\@secondoftwo}%
\providecommand \bibfield [0]{\@secondoftwo}%
\providecommand \translation [1]{[#1]}%
\providecommand \BibitemOpen[0]{}%
\providecommand \bibitemStop [0]{}%
\providecommand \bibitemNoStop [0]{.\EOS\space}%
\providecommand \EOS [0]{\spacefactor3000\relax}%
\providecommand \BibitemShut [1]{\csname bibitem#1\endcsname}%
\bibitem{Florkowski:2010zz}%
  \BibitemOpen
  \bibfield{author}{%
  \bibinfo {author} {\bibfnamefont{W.}~\bibnamefont{Florkowski}},\ }%
  \emph{\bibinfo {title} {{Phenomenology of Ultra-Relativistic Heavy-Ion
  Collisions}}}\ (\bibinfo {publisher} {World Scientific Publishing Company,
  Singapore},\ \bibinfo {year} {2010})%
  \bibAnnoteFile{NoStop}{Florkowski:2010zz}%
\bibitem{Heinz:2013th}%
  \BibitemOpen
  \bibfield{author}{%
  \bibinfo {author} {\bibfnamefont{U.}~\bibnamefont{Heinz}}\ and\ \bibinfo
  {author} {\bibfnamefont{R.}~\bibnamefont{Snellings}},\ }%
  \bibfield{journal}{%
  \Doi{10.1146/annurev-nucl-102212-170540}{\bibinfo {journal}
  {Ann.Rev.Nucl.Part.Sci.}}\ }%
  \textbf{\bibinfo {volume} {63}},\ \bibinfo {pages} {123} (\bibinfo {year}
  {2013})%
  \bibAnnoteFile{NoStop}{Heinz:2013th}%
\bibitem{Gale:2013da}%
  \BibitemOpen
  \bibfield{author}{%
  \bibinfo {author} {\bibfnamefont{C.}~\bibnamefont{Gale}}, \bibinfo {author}
  {\bibfnamefont{S.}~\bibnamefont{Jeon}},\ and\ \bibinfo {author}
  {\bibfnamefont{B.}~\bibnamefont{Schenke}},\ }%
  \bibfield{journal}{%
  \Doi{10.1142/S0217751X13400113}{\bibinfo {journal} {Int.J.Mod.Phys.}}\ }%
  \textbf{\bibinfo {volume} {A28}},\ \bibinfo {pages} {1340011} (\bibinfo
  {year} {2013})%
  \bibAnnoteFile{NoStop}{Gale:2013da}%
\bibitem{Luzum:2013yya}%
  \BibitemOpen
  \bibfield{author}{%
  \bibinfo {author} {\bibfnamefont{M.}~\bibnamefont{Luzum}}\ and\ \bibinfo
  {author} {\bibfnamefont{H.}~\bibnamefont{Petersen}}}%
   (\bibinfo {year} {2013}),\
  \Eprint{http://arxiv.org/abs/1312.5503}{arXiv:1312.5503 [nucl-th]}%
  \bibAnnoteFile{NoStop}{Luzum:2013yya}%
\bibitem{Ollitrault:1992}%
  \BibitemOpen
  \bibfield{author}{%
  \bibinfo {author} {\bibfnamefont{J.~Y.}\ \bibnamefont{Ollitrault}},\ }%
  \bibfield{journal}{%
  \bibinfo {journal} {Phys. Rev.}\ }%
  \textbf{\bibinfo {volume} {D46}},\ \bibinfo {pages} {229} (\bibinfo {year}
  {1992})%
  \bibAnnoteFile{NoStop}{Ollitrault:1992}%
\bibitem{Alver:2006wh}%
  \BibitemOpen
  \bibfield{author}{%
  \bibinfo {author} {\bibfnamefont{B.}~\bibnamefont{Alver}} \emph{et~al.}
  (\bibinfo {collaboration} {PHOBOS Collaboration}),\ }%
  \bibfield{journal}{%
  \Doi{10.1103/PhysRevLett.98.242302}{\bibinfo {journal} {Phys. Rev. Lett.}}\
  }%
  \textbf{\bibinfo {volume} {98}},\ \bibinfo {pages} {242302} (\bibinfo {year}
  {2007})%
  \bibAnnoteFile{NoStop}{Alver:2006wh}%
\bibitem{Alver:2010gr}%
  \BibitemOpen
  \bibfield{author}{%
  \bibinfo {author} {\bibfnamefont{B.}~\bibnamefont{Alver}}\ and\ \bibinfo
  {author} {\bibfnamefont{G.}~\bibnamefont{Roland}},\ }%
  \bibfield{journal}{%
  \Doi{10.1103/PhysRevC.81.054905}{\bibinfo {journal} {Phys. Rev.}}\ }%
  \textbf{\bibinfo {volume} {C81}},\ \bibinfo {pages} {054905} (\bibinfo {year}
  {2010})%
  \bibAnnoteFile{NoStop}{Alver:2010gr}%
\bibitem{Petersen:2010cw}%
  \BibitemOpen
  \bibfield{author}{%
  \bibinfo {author} {\bibfnamefont{H.}~\bibnamefont{Petersen}}, \bibinfo
  {author} {\bibfnamefont{G.-Y.}\ \bibnamefont{Qin}}, \bibinfo {author}
  {\bibfnamefont{S.~A.}\ \bibnamefont{Bass}},\ and\ \bibinfo {author}
  {\bibfnamefont{B.}~\bibnamefont{Muller}},\ }%
  \bibfield{journal}{%
  \Doi{10.1103/PhysRevC.82.041901}{\bibinfo {journal} {Phys. Rev.}}\ }%
  \textbf{\bibinfo {volume} {C82}},\ \bibinfo {pages} {041901} (\bibinfo {year}
  {2010})%
  \bibAnnoteFile{NoStop}{Petersen:2010cw}%
\bibitem{Alver:2010dn}%
  \BibitemOpen
  \bibfield{author}{%
  \bibinfo {author} {\bibfnamefont{B.~H.}\ \bibnamefont{Alver}}, \bibinfo
  {author} {\bibfnamefont{C.}~\bibnamefont{Gombeaud}}, \bibinfo {author}
  {\bibfnamefont{M.}~\bibnamefont{Luzum}},\ and\ \bibinfo {author}
  {\bibfnamefont{J.-Y.}\ \bibnamefont{Ollitrault}},\ }%
  \bibfield{journal}{%
  \Doi{10.1103/PhysRevC.82.034913}{\bibinfo {journal} {Phys. Rev.}}\ }%
  \textbf{\bibinfo {volume} {C82}},\ \bibinfo {pages} {034913} (\bibinfo {year}
  {2010})%
  \bibAnnoteFile{NoStop}{Alver:2010dn}%
\bibitem{Wiedemann:1999qn}%
  \BibitemOpen
  \bibfield{author}{%
  \bibinfo {author} {\bibfnamefont{U.~A.}\ \bibnamefont{Wiedemann}}\ and\
  \bibinfo {author} {\bibfnamefont{U.~W.}\ \bibnamefont{Heinz}},\ }%
  \bibfield{journal}{%
  \Doi{10.1016/S0370-1573(99)00032-0}{\bibinfo {journal} {Phys. Rept.}}\ }%
  \textbf{\bibinfo {volume} {319}},\ \bibinfo {pages} {145} (\bibinfo {year}
  {1999})%
  \bibAnnoteFile{NoStop}{Wiedemann:1999qn}%
\bibitem{Heinz:1999rw}%
  \BibitemOpen
  \bibfield{author}{%
  \bibinfo {author} {\bibfnamefont{U.~W.}\ \bibnamefont{Heinz}}\ and\ \bibinfo
  {author} {\bibfnamefont{B.~V.}\ \bibnamefont{Jacak}},\ }%
  \bibfield{journal}{%
  \Doi{10.1146/annurev.nucl.49.1.529}{\bibinfo {journal} {Ann. Rev. Nucl. Part.
  Sci.}}\ }%
  \textbf{\bibinfo {volume} {49}},\ \bibinfo {pages} {529} (\bibinfo {year}
  {1999})%
  \bibAnnoteFile{NoStop}{Heinz:1999rw}%
\bibitem{Lisa:2005dd}%
  \BibitemOpen
  \bibfield{author}{%
  \bibinfo {author} {\bibfnamefont{M.~A.}\ \bibnamefont{Lisa}}, \bibinfo
  {author} {\bibfnamefont{S.}~\bibnamefont{Pratt}}, \bibinfo {author}
  {\bibfnamefont{R.}~\bibnamefont{Soltz}},\ and\ \bibinfo {author}
  {\bibfnamefont{U.}~\bibnamefont{Wiedemann}},\ }%
  \bibfield{journal}{%
  \bibinfo {journal} {Ann. Rev. Nucl. Part. Sci.}\ }%
  \textbf{\bibinfo {volume} {55}},\ \bibinfo {pages} {357} (\bibinfo {year}
  {2005})%
  \bibAnnoteFile{NoStop}{Lisa:2005dd}%
\bibitem{Lisa:2011na}%
  \BibitemOpen
  \bibfield{author}{%
  \bibinfo {author} {\bibfnamefont{M.}~\bibnamefont{Lisa}}, \bibinfo {author}
  {\bibfnamefont{E.}~\bibnamefont{Frodermann}}, \bibinfo {author}
  {\bibfnamefont{G.}~\bibnamefont{Graef}}, \bibinfo {author}
  {\bibfnamefont{M.}~\bibnamefont{Mitrovski}}, \bibinfo {author}
  {\bibfnamefont{E.}~\bibnamefont{Mount}}, \emph{et~al.},\ }%
  \bibfield{journal}{%
  \Doi{10.1088/1367-2630/13/6/065006}{\bibinfo {journal} {New J.Phys.}}\ }%
  \textbf{\bibinfo {volume} {13}},\ \bibinfo {pages} {065006} (\bibinfo {year}
  {2011})%
  \bibAnnoteFile{NoStop}{Lisa:2011na}%
\bibitem{Akkelin:1995gh}%
  \BibitemOpen
  \bibfield{author}{%
  \bibinfo {author} {\bibfnamefont{S.}~\bibnamefont{Akkelin}}\ and\ \bibinfo
  {author} {\bibfnamefont{Y.}~\bibnamefont{Sinyukov}},\ }%
  \bibfield{journal}{%
  \Doi{10.1016/0370-2693(95)00765-D}{\bibinfo {journal} {Phys. Lett.}}\ }%
  \textbf{\bibinfo {volume} {B356}},\ \bibinfo {pages} {525} (\bibinfo {year}
  {1995})%
  \bibAnnoteFile{NoStop}{Akkelin:1995gh}%
\bibitem{Abelev:2009tp}%
  \BibitemOpen
  \bibfield{author}{%
  \bibinfo {author} {\bibfnamefont{B.}~\bibnamefont{Abelev}} \emph{et~al.}
  (\bibinfo {collaboration} {STAR Collaboration}),\ }%
  \bibfield{journal}{%
  \Doi{10.1103/PhysRevC.80.024905}{\bibinfo {journal} {Phys. Rev.}}\ }%
  \textbf{\bibinfo {volume} {C80}},\ \bibinfo {pages} {024905} (\bibinfo {year}
  {2009})%
  \bibAnnoteFile{NoStop}{Abelev:2009tp}%
\bibitem{Adams:2004yc}%
  \BibitemOpen
  \bibfield{author}{%
  \bibinfo {author} {\bibfnamefont{J.}~\bibnamefont{Adams}} \emph{et~al.}
  (\bibinfo {collaboration} {STAR Collaboration}),\ }%
  \bibfield{journal}{%
  \bibinfo {journal} {Phys. Rev.}\ }%
  \textbf{\bibinfo {volume} {C71}},\ \bibinfo {pages} {044906} (\bibinfo {year}
  {2005})%
  \bibAnnoteFile{NoStop}{Adams:2004yc}%
\bibitem{Aamodt:2011mr}%
  \BibitemOpen
  \bibfield{author}{%
  \bibinfo {author} {\bibfnamefont{K.}~\bibnamefont{Aamodt}} \emph{et~al.}
  (\bibinfo {collaboration} {ALICE Collaboration}),\ }%
  \bibfield{journal}{%
  \Doi{10.1016/j.physletb.2010.12.053}{\bibinfo {journal} {Phys. Lett.}}\ }%
  \textbf{\bibinfo {volume} {B696}},\ \bibinfo {pages} {328} (\bibinfo {year}
  {2011})%
  \bibAnnoteFile{NoStop}{Aamodt:2011mr}%
\bibitem{Broniowski:2008vp}%
  \BibitemOpen
  \bibfield{author}{%
  \bibinfo {author} {\bibfnamefont{W.}~\bibnamefont{Broniowski}}, \bibinfo
  {author} {\bibfnamefont{M.}~\bibnamefont{Chojnacki}}, \bibinfo {author}
  {\bibfnamefont{W.}~\bibnamefont{Florkowski}},\ and\ \bibinfo {author}
  {\bibfnamefont{A.}~\bibnamefont{Kisiel}},\ }%
  \bibfield{journal}{%
  \Doi{10.1103/PhysRevLett.101.022301}{\bibinfo {journal} {Phys. Rev. Lett.}}\
  }%
  \textbf{\bibinfo {volume} {101}},\ \bibinfo {pages} {022301} (\bibinfo {year}
  {2008})%
  \bibAnnoteFile{NoStop}{Broniowski:2008vp}%
\bibitem{Pratt:2008qv}%
  \BibitemOpen
  \bibfield{author}{%
  \bibinfo {author} {\bibfnamefont{S.}~\bibnamefont{Pratt}},\ }%
  \bibfield{journal}{%
  \Doi{10.1103/PhysRevLett.102.232301}{\bibinfo {journal} {Phys. Rev. Lett.}}\
  }%
  \textbf{\bibinfo {volume} {102}},\ \bibinfo {pages} {232301} (\bibinfo {year}
  {2009})%
  \bibAnnoteFile{NoStop}{Pratt:2008qv}%
\bibitem{Bozek:2010er}%
  \BibitemOpen
  \bibfield{author}{%
  \bibinfo {author} {\bibfnamefont{P.}~\bibnamefont{Bo\.zek}},\ }%
  \bibfield{journal}{%
  \Doi{10.1103/PhysRevC.83.044910}{\bibinfo {journal} {Phys. Rev.}}\ }%
  \textbf{\bibinfo {volume} {C83}},\ \bibinfo {pages} {044910} (\bibinfo {year}
  {2011})%
  \bibAnnoteFile{NoStop}{Bozek:2010er}%
\bibitem{Karpenko:2012yf}%
  \BibitemOpen
  \bibfield{author}{%
  \bibinfo {author} {\bibfnamefont{I.}~\bibnamefont{Karpenko}}, \bibinfo
  {author} {\bibfnamefont{Y.}~\bibnamefont{Sinyukov}},\ and\ \bibinfo {author}
  {\bibfnamefont{K.}~\bibnamefont{Werner}},\ }%
  \bibfield{journal}{%
  \Doi{10.1103/PhysRevC.87.024914}{\bibinfo {journal} {Phys. Rev.}}\ }%
  \textbf{\bibinfo {volume} {C87}},\ \bibinfo {pages} {024914} (\bibinfo {year}
  {2013})%
  \bibAnnoteFile{NoStop}{Karpenko:2012yf}%
\bibitem{Voloshin:1995mc}%
  \BibitemOpen
  \bibfield{author}{%
  \bibinfo {author} {\bibfnamefont{S.}~\bibnamefont{Voloshin}}\ and\ \bibinfo
  {author} {\bibfnamefont{W.}~\bibnamefont{Cleland}},\ }%
  \bibfield{journal}{%
  \Doi{10.1103/PhysRevC.53.896}{\bibinfo {journal} {Phys.Rev.}}\ }%
  \textbf{\bibinfo {volume} {C53}},\ \bibinfo {pages} {896} (\bibinfo {year}
  {1996})%
  \bibAnnoteFile{NoStop}{Voloshin:1995mc}%
\bibitem{Wiedemann:1997cr}%
  \BibitemOpen
  \bibfield{author}{%
  \bibinfo {author} {\bibfnamefont{U.~A.}\ \bibnamefont{Wiedemann}},\ }%
  \bibfield{journal}{%
  \Doi{10.1103/PhysRevC.57.266}{\bibinfo {journal} {Phys.Rev.}}\ }%
  \textbf{\bibinfo {volume} {C57}},\ \bibinfo {pages} {266} (\bibinfo {year}
  {1998})%
  \bibAnnoteFile{NoStop}{Wiedemann:1997cr}%
\bibitem{Lisa:2000ip}%
  \BibitemOpen
  \bibfield{author}{%
  \bibinfo {author} {\bibfnamefont{M.~A.}\ \bibnamefont{Lisa}}, \bibinfo
  {author} {\bibfnamefont{U.~W.}\ \bibnamefont{Heinz}},\ and\ \bibinfo {author}
  {\bibfnamefont{U.~A.}\ \bibnamefont{Wiedemann}},\ }%
  \bibfield{journal}{%
  \Doi{10.1016/S0370-2693(00)00952-7}{\bibinfo {journal} {Phys. Lett.}}\ }%
  \textbf{\bibinfo {volume} {B489}},\ \bibinfo {pages} {287} (\bibinfo {year}
  {2000})%
  \bibAnnoteFile{NoStop}{Lisa:2000ip}%
\bibitem{Heinz:2002sq}%
  \BibitemOpen
  \bibfield{author}{%
  \bibinfo {author} {\bibfnamefont{U.~W.}\ \bibnamefont{Heinz}}\ and\ \bibinfo
  {author} {\bibfnamefont{P.~F.}\ \bibnamefont{Kolb}},\ }%
  \bibfield{journal}{%
  \Doi{10.1016/S0370-2693(02)02372-9}{\bibinfo {journal} {Phys.Lett.}}\ }%
  \textbf{\bibinfo {volume} {B542}},\ \bibinfo {pages} {216} (\bibinfo {year}
  {2002})%
  \bibAnnoteFile{NoStop}{Heinz:2002sq}%
\bibitem{Heinz:2002au}%
  \BibitemOpen
  \bibfield{author}{%
  \bibinfo {author} {\bibfnamefont{U.~W.}\ \bibnamefont{Heinz}}, \bibinfo
  {author} {\bibfnamefont{A.}~\bibnamefont{Hummel}}, \bibinfo {author}
  {\bibfnamefont{M.}~\bibnamefont{Lisa}},\ and\ \bibinfo {author}
  {\bibfnamefont{U.}~\bibnamefont{Wiedemann}},\ }%
  \bibfield{journal}{%
  \Doi{10.1103/PhysRevC.66.044903}{\bibinfo {journal} {Phys.Rev.}}\ }%
  \textbf{\bibinfo {volume} {C66}},\ \bibinfo {pages} {044903} (\bibinfo {year}
  {2002})%
  \bibAnnoteFile{NoStop}{Heinz:2002au}%
\bibitem{Lisa:2000xj}%
  \BibitemOpen
  \bibfield{author}{%
  \bibinfo {author} {\bibfnamefont{M.}~\bibnamefont{Lisa}} \emph{et~al.}
  (\bibinfo {collaboration} {E895 Collaboration}),\ }%
  \bibfield{journal}{%
  \Doi{10.1016/S0370-2693(00)01280-6}{\bibinfo {journal} {Phys.Lett.}}\ }%
  \textbf{\bibinfo {volume} {B496}},\ \bibinfo {pages} {1} (\bibinfo {year}
  {2000})%
  \bibAnnoteFile{NoStop}{Lisa:2000xj}%
\bibitem{Adams:2003ra}%
  \BibitemOpen
  \bibfield{author}{%
  \bibinfo {author} {\bibfnamefont{J.}~\bibnamefont{Adams}} \emph{et~al.}
  (\bibinfo {collaboration} {STAR Collaboration}),\ }%
  \bibfield{journal}{%
  \Doi{10.1103/PhysRevLett.93.012301}{\bibinfo {journal} {Phys.Rev.Lett.}}\ }%
  \textbf{\bibinfo {volume} {93}},\ \bibinfo {pages} {012301} (\bibinfo {year}
  {2004})%
  \bibAnnoteFile{NoStop}{Adams:2003ra}%
\bibitem{Adamova:2008hs}%
  \BibitemOpen
  \bibfield{author}{%
  \bibinfo {author} {\bibfnamefont{D.}~\bibnamefont{Adamova}} \emph{et~al.}
  (\bibinfo {collaboration} {CERES Collaboration}),\ }%
  \bibfield{journal}{%
  \Doi{10.1103/PhysRevC.78.064901}{\bibinfo {journal} {Phys.Rev.}}\ }%
  \textbf{\bibinfo {volume} {C78}},\ \bibinfo {pages} {064901} (\bibinfo {year}
  {2008})%
  \bibAnnoteFile{NoStop}{Adamova:2008hs}%
\bibitem{Gramling:2012xqa}%
  \BibitemOpen
  \bibinfo {author} {\bibfnamefont{J.}~\bibnamefont{Gramling,\ }}%
  \bibfield{journal}{%
  \bibinfo {journal} {CERN-THESIS-2012-088}}%
   (\bibinfo {year} {2012})%
  \bibAnnoteFile{NoStop}{Gramling:2012xqa}%
\bibitem{logginswpcf}%
  \BibitemOpen
\bibfield{author}{%
    }%
  \bibfield{author}{%
  \bibinfo {author} {\bibfnamefont{V.}~\bibnamefont{Loggins}} (\bibinfo
  {collaboration} {ALICE Collaboration}),\ }%
  \bibfield{journal}{%
  \bibinfo {journal} {talk given at the Workshop on Correlations and
  Femtoscopy, Acireale, Nov 5-8}}%
   (\bibinfo {year} {2013})%
  \bibAnnoteFile{NoStop}{logginswpcf}%
\bibitem{Voloshin:2011mg}%
  \BibitemOpen
  \bibfield{author}{%
  \bibinfo {author} {\bibfnamefont{S.~A.}\ \bibnamefont{Voloshin}},\ }%
  \bibfield{journal}{%
  \Doi{10.1088/0954-3899/38/12/124097}{\bibinfo {journal} {J.Phys.}}\ }%
  \textbf{\bibinfo {volume} {G38}},\ \bibinfo {pages} {124097} (\bibinfo {year}
  {2011})%
  \bibAnnoteFile{NoStop}{Voloshin:2011mg}%
\bibitem{Niida:2013lia}%
  \BibitemOpen
  \bibfield{author}{%
  \bibinfo {author} {\bibfnamefont{T.}~\bibnamefont{Niida}} (\bibinfo
  {collaboration} {PHENIX Collaboration}),\ }%
  \bibfield{journal}{%
  \Doi{10.1016/j.nuclphysa.2013.02.043}{\bibinfo {journal} {Nucl.Phys.}}\ }%
  \textbf{\bibinfo {volume} {A904-905}},\ \bibinfo {pages} {439c} (\bibinfo
  {year} {2013})%
  \bibAnnoteFile{NoStop}{Niida:2013lia}%
\bibitem{niidawpcf}%
  \BibitemOpen
  \bibfield{author}{%
  \bibinfo {author} {\bibfnamefont{T.}~\bibnamefont{Niida}} (\bibinfo
  {collaboration} {PHENIX Collaboration}),\ }%
  \bibfield{journal}{%
  \bibinfo {journal} {talk given at the Workshop on Correlations and
  Femtoscopy, Acireale, Nov 5-8}}%
   (\bibinfo {year} {2013})%
  \bibAnnoteFile{NoStop}{niidawpcf}%
\bibitem{Plumberg:2013nga}%
  \BibitemOpen
  \bibfield{author}{%
  \bibinfo {author} {\bibfnamefont{C.~J.}\ \bibnamefont{Plumberg}}, \bibinfo
  {author} {\bibfnamefont{C.}~\bibnamefont{Shen}},\ and\ \bibinfo {author}
  {\bibfnamefont{U.~W.}\ \bibnamefont{Heinz}},\ }%
  \bibfield{journal}{%
  \Doi{10.1103/PhysRevC.88.044914}{\bibinfo {journal} {Phys.Rev.}}\ }%
  \textbf{\bibinfo {volume} {C88}},\ \bibinfo {pages} {044914} (\bibinfo {year}
  {2013})%
  \bibAnnoteFile{NoStop}{Plumberg:2013nga}%
\bibitem{Frodermann:2007ab}%
  \BibitemOpen
  \bibfield{author}{%
  \bibinfo {author} {\bibfnamefont{E.}~\bibnamefont{Frodermann}}, \bibinfo
  {author} {\bibfnamefont{R.}~\bibnamefont{Chatterjee}},\ and\ \bibinfo
  {author} {\bibfnamefont{U.}~\bibnamefont{Heinz}},\ }%
  \bibfield{journal}{%
  \Doi{10.1088/0954-3899/34/11/002}{\bibinfo {journal} {J.Phys.}}\ }%
  \textbf{\bibinfo {volume} {G34}},\ \bibinfo {pages} {2249} (\bibinfo {year}
  {2007})%
  \bibAnnoteFile{NoStop}{Frodermann:2007ab}%
\bibitem{Kisiel:2008ws}%
  \BibitemOpen
  \bibfield{author}{%
  \bibinfo {author} {\bibfnamefont{A.}~\bibnamefont{Kisiel}}, \bibinfo {author}
  {\bibfnamefont{W.}~\bibnamefont{Broniowski}}, \bibinfo {author}
  {\bibfnamefont{M.}~\bibnamefont{Chojnacki}},\ and\ \bibinfo {author}
  {\bibfnamefont{W.}~\bibnamefont{Florkowski}},\ }%
  \bibfield{journal}{%
  \Doi{10.1103/PhysRevC.79.014902}{\bibinfo {journal} {Phys. Rev.}}\ }%
  \textbf{\bibinfo {volume} {C79}},\ \bibinfo {pages} {014902} (\bibinfo {year}
  {2009})%
  \bibAnnoteFile{NoStop}{Kisiel:2008ws}%
\bibitem{Socolowski:2004hw}%
  \BibitemOpen
  \bibfield{author}{%
  \bibinfo {author} {\bibfnamefont{J.}~\bibnamefont{Socolowski},
  \bibfnamefont{O.}}, \bibinfo {author}
  {\bibfnamefont{F.}~\bibnamefont{Grassi}}, \bibinfo {author}
  {\bibfnamefont{Y.}~\bibnamefont{Hama}},\ and\ \bibinfo {author}
  {\bibfnamefont{T.}~\bibnamefont{Kodama}},\ }%
  \bibfield{journal}{%
  \Doi{10.1103/PhysRevLett.93.182301}{\bibinfo {journal} {Phys.Rev.Lett.}}\ }%
  \textbf{\bibinfo {volume} {93}},\ \bibinfo {pages} {182301} (\bibinfo {year}
  {2004})%
  \bibAnnoteFile{NoStop}{Socolowski:2004hw}%
\bibitem{Bozek:2012hy}%
  \BibitemOpen
  \bibfield{author}{%
  \bibinfo {author} {\bibfnamefont{P.}~\bibnamefont{Bo\.zek}},\ }%
  \bibfield{journal}{%
  \Doi{10.1016/j.physletb.2012.09.040}{\bibinfo {journal} {Phys. Lett.}}\ }%
  \textbf{\bibinfo {volume} {B717}},\ \bibinfo {pages} {287} (\bibinfo {year}
  {2012})%
  \bibAnnoteFile{NoStop}{Bozek:2012hy}%
\bibitem{Schenke:2010rr}%
  \BibitemOpen
  \bibfield{author}{%
  \bibinfo {author} {\bibfnamefont{B.}~\bibnamefont{Schenke}}, \bibinfo
  {author} {\bibfnamefont{S.}~\bibnamefont{Jeon}},\ and\ \bibinfo {author}
  {\bibfnamefont{C.}~\bibnamefont{Gale}},\ }%
  \bibfield{journal}{%
  \Doi{10.1103/PhysRevLett.106.042301}{\bibinfo {journal} {Phys. Rev. Lett.}}\
  }%
  \textbf{\bibinfo {volume} {106}},\ \bibinfo {pages} {042301} (\bibinfo {year}
  {2011})%
  \bibAnnoteFile{NoStop}{Schenke:2010rr}%
\bibitem{Li:2012ta}%
  \BibitemOpen
  \bibfield{author}{%
  \bibinfo {author} {\bibfnamefont{Q.}~\bibnamefont{Li}}, \bibinfo {author}
  {\bibfnamefont{G.}~\bibnamefont{Graf}},\ and\ \bibinfo {author}
  {\bibfnamefont{M.}~\bibnamefont{Bleicher}},\ }%
  \bibfield{journal}{%
  \Doi{10.1103/PhysRevC.85.034908}{\bibinfo {journal} {Phys.Rev.}}\ }%
  \textbf{\bibinfo {volume} {C85}},\ \bibinfo {pages} {034908} (\bibinfo {year}
  {2012})%
  \bibAnnoteFile{NoStop}{Li:2012ta}%
\bibitem{Graef:2013wta}%
  \BibitemOpen
  \bibfield{author}{%
  \bibinfo {author} {\bibfnamefont{G.}~\bibnamefont{Graef}}, \bibinfo {author}
  {\bibfnamefont{M.}~\bibnamefont{Lisa}},\ and\ \bibinfo {author}
  {\bibfnamefont{M.}~\bibnamefont{Bleicher}},\ }%
  \bibfield{journal}{%
  \Doi{10.1103/PhysRevC.89.014903}{\bibinfo {journal} {Phys.Rev.}}\ }%
  \textbf{\bibinfo {volume} {C89}},\ \bibinfo {pages} {014903} (\bibinfo {year}
  {2014})%
  \bibAnnoteFile{NoStop}{Graef:2013wta}%
\bibitem{Bozek:2011ua}%
  \BibitemOpen
  \bibfield{author}{%
  \bibinfo {author} {\bibfnamefont{P.}~\bibnamefont{Bo\.zek}},\ }%
  \bibfield{journal}{%
  \Doi{10.1103/PhysRevC.85.034901}{\bibinfo {journal} {Phys. Rev.}}\ }%
  \textbf{\bibinfo {volume} {C85}},\ \bibinfo {pages} {034901} (\bibinfo {year}
  {2012})%
  \bibAnnoteFile{NoStop}{Bozek:2011ua}%
\bibitem{Bozek:2009dw}%
  \BibitemOpen
  \bibfield{author}{%
  \bibinfo {author} {\bibfnamefont{P.}~\bibnamefont{Bo\.zek}},\ }%
  \bibfield{journal}{%
  \bibinfo {journal} {Phys. Rev.}\ }%
  \textbf{\bibinfo {volume} {C81}},\ \bibinfo {pages} {034909} (\bibinfo {year}
  {2010})%
  \bibAnnoteFile{NoStop}{Bozek:2009dw}%
\bibitem{Rybczynski:2013yba}%
  \BibitemOpen
  \bibfield{author}{%
  \bibinfo {author} {\bibfnamefont{M.}~\bibnamefont{Rybczynski}}, \bibinfo
  {author} {\bibfnamefont{G.}~\bibnamefont{Stefanek}}, \bibinfo {author}
  {\bibfnamefont{W.}~\bibnamefont{Broniowski}},\ and\ \bibinfo {author}
  {\bibfnamefont{P.}~\bibnamefont{Bozek}}}%
   (\bibinfo {year} {2013}),\
  \Eprint{http://arxiv.org/abs/1310.5475}{arXiv:1310.5475 [nucl-th]}%
  \bibAnnoteFile{NoStop}{Rybczynski:2013yba}%
\bibitem{Bozek:2011if}%
  \BibitemOpen
  \bibfield{author}{%
  \bibinfo {author} {\bibfnamefont{P.}~\bibnamefont{Bo\.zek}},\ }%
  \bibfield{journal}{%
  \bibinfo {journal} {Phys. Rev.}\ }%
  \textbf{\bibinfo {volume} {C85}},\ \bibinfo {pages} {014911} (\bibinfo {year}
  {2012})%
  \bibAnnoteFile{NoStop}{Bozek:2011if}%
\bibitem{Bozek:2012fw}%
  \BibitemOpen
  \bibfield{author}{%
  \bibinfo {author} {\bibfnamefont{P.}~\bibnamefont{Bo\.zek}}\ and\ \bibinfo
  {author} {\bibfnamefont{W.}~\bibnamefont{Broniowski}},\ }%
  \bibfield{journal}{%
  \bibinfo {journal} {Phys. Rev.}\ }%
  \textbf{\bibinfo {volume} {C85}},\ \bibinfo {pages} {044910} (\bibinfo {year}
  {2012})%
  \bibAnnoteFile{NoStop}{Bozek:2012fw}%
\bibitem{Chojnacki:2011hb}%
  \BibitemOpen
  \bibfield{author}{%
  \bibinfo {author} {\bibfnamefont{M.}~\bibnamefont{Chojnacki}}, \bibinfo
  {author} {\bibfnamefont{A.}~\bibnamefont{Kisiel}}, \bibinfo {author}
  {\bibfnamefont{W.}~\bibnamefont{Florkowski}},\ and\ \bibinfo {author}
  {\bibfnamefont{W.}~\bibnamefont{Broniowski}},\ }%
  \bibfield{journal}{%
  \Doi{10.1016/j.cpc.2011.11.018}{\bibinfo {journal} {Comput. Phys. Commun.}}\
  }%
  \textbf{\bibinfo {volume} {183}},\ \bibinfo {pages} {746} (\bibinfo {year}
  {2012})%
  \bibAnnoteFile{NoStop}{Chojnacki:2011hb}%
\bibitem{Retiere:2003kf}%
  \BibitemOpen
  \bibfield{author}{%
  \bibinfo {author} {\bibfnamefont{F.}~\bibnamefont{Retiere}}\ and\ \bibinfo
  {author} {\bibfnamefont{M.~A.}\ \bibnamefont{Lisa}},\ }%
  \bibfield{journal}{%
  \Doi{10.1103/PhysRevC.70.044907}{\bibinfo {journal} {Phys. Rev.}}\ }%
  \textbf{\bibinfo {volume} {C70}},\ \bibinfo {pages} {044907} (\bibinfo {year}
  {2004})%
  \bibAnnoteFile{NoStop}{Retiere:2003kf}%
\bibitem{Bowler:1991vx}%
  \BibitemOpen
  \bibfield{author}{%
  \bibinfo {author} {\bibfnamefont{M.~G.}\ \bibnamefont{Bowler}},\ }%
  \bibfield{journal}{%
  \Doi{10.1016/0370-2693(91)91541-3}{\bibinfo {journal} {Phys. Lett.}}\ }%
  \textbf{\bibinfo {volume} {B270}},\ \bibinfo {pages} {69} (\bibinfo {year}
  {1991})%
  \bibAnnoteFile{NoStop}{Bowler:1991vx}%
\bibitem{Sinyukov:1998fc}%
  \BibitemOpen
  \bibfield{author}{%
  \bibinfo {author} {\bibfnamefont{Y.}~\bibnamefont{Sinyukov}}, \bibinfo
  {author} {\bibfnamefont{R.}~\bibnamefont{Lednicky}}, \bibinfo {author}
  {\bibfnamefont{S.~V.}\ \bibnamefont{Akkelin}}, \bibinfo {author}
  {\bibfnamefont{J.}~\bibnamefont{Pluta}},\ and\ \bibinfo {author}
  {\bibfnamefont{B.}~\bibnamefont{Erazmus}},\ }%
  \bibfield{journal}{%
  \Doi{10.1016/S0370-2693(98)00653-4}{\bibinfo {journal} {Phys. Lett.}}\ }%
  \textbf{\bibinfo {volume} {B432}},\ \bibinfo {pages} {248} (\bibinfo {year}
  {1998})%
  \bibAnnoteFile{NoStop}{Sinyukov:1998fc}%
\bibitem{Bertsch:1989vn}%
  \BibitemOpen
  \bibfield{author}{%
  \bibinfo {author} {\bibfnamefont{G.~F.}\ \bibnamefont{Bertsch}},\ }%
  \bibfield{journal}{%
  \bibinfo {journal} {Nucl. Phys.}\ }%
  \textbf{\bibinfo {volume} {A498}},\ \bibinfo {pages} {173c} (\bibinfo {year}
  {1989})%
  \bibAnnoteFile{NoStop}{Bertsch:1989vn}%
\bibitem{Pratt:1986cc}%
  \BibitemOpen
  \bibfield{author}{%
  \bibinfo {author} {\bibfnamefont{S.}~\bibnamefont{Pratt}},\ }%
  \bibfield{journal}{%
  \Doi{10.1103/PhysRevD.33.1314}{\bibinfo {journal} {Phys. Rev.}}\ }%
  \textbf{\bibinfo {volume} {D33}},\ \bibinfo {pages} {1314} (\bibinfo {year}
  {1986})%
  \bibAnnoteFile{NoStop}{Pratt:1986cc}%
\end{thebibliography}%

\end{document}